\documentclass[10pt,amsmath,amssymb,aps,prb,twocolumn,]{revtex4-2}

\usepackage{graphicx} 
\usepackage{color} 

\usepackage{xcolor}      % Extended colors
\usepackage{dcolumn}     % Align table columns on decimal point
\usepackage{bm}          % Bold math symbols
\usepackage{float}       % To better control float positions
\begin{document}

\preprint{APS/123-QED}

\title{A Matrix Quantum Kinetic Treatment of Impact Ionization in Avalanche Photodiodes}% Force line breaks with \\

\author{Sheikh Z. Ahmed}
\thanks{These authors contributed equally to this work.}
\affiliation{Department of Electrical and Computer Engineering, University of Virginia, Charlottesville, Virginia 22904, USA}%Lines break automatically or can be forced with \\
\author{Shafat Shahnewaz}
\thanks{These authors contributed equally to this work.}
\affiliation{Department of Electrical and Computer Engineering, University of Virginia, Charlottesville, Virginia 22904, USA}
\author{Samiran Ganguly}%
\affiliation{Department of Electrical and Computer Engineering, University of Virginia, Charlottesville, Virginia 22904, USA}
\author{Joe C Campbell}
\affiliation{Department of Electrical and Computer Engineering, University of Virginia, Charlottesville, Virginia 22904, USA}
\author{Avik W. Ghosh}
\email{ag7rq@virginia.edu}
\affiliation{Department of Electrical and Computer Engineering, University of Virginia, Charlottesville, Virginia 22904, USA}
\affiliation{Department of Physics, University of Virginia, Charlottesville, Virginia 22904, USA}

\date{\today}% It is always \today, today,

\begin{abstract}
 Matrix based quantum kinetic simulations have been widely used for the predictive modeling of electronic devices. Inelastic scattering from phonons and electrons are typically treated as higher order processes in these treatments, captured using mean-field approximations. Carrier multiplication in Avalanche Photodiodes (APDs), however, relies entirely on strongly inelastic impact ionization, making electron-electron scattering the dominant term requiring a rigorous, microscopic treatment. We go well beyond the conventional Born approximation for scattering to develop a matrix-based quantum kinetic theory for impact ionization, involving products of multiple Green's functions.  Using a model semiconductor in a reverse-biased p-i-n configuration, we show how its calculated non-equilibrium charge distributions show multiplication at dead-space values consistent with energy-momentum conservation. Our matrix approach can be readily generalized to more sophisticated atomistic Hamiltonians, setting the stage for a fully predictive, `first principles' theory of APDs. 
\end{abstract}

\maketitle

%\tableofcontents

\section{\label{sec:Introduction}Introduction}
Avalanche photodiodes (APD) are commercially employed for a wide range of applications, ranging from silicon photonics to 
light imaging, detection and ranging (LIDAR) to single photon sensing and night vision \cite{Tosi, CAMPBELL2008221, bertone2007avalanche, Mitra2006, Williams2017, nada2020high, Pasquinelli2020, Thomson_2016}. These applications capitalize on the highly efficient photodetection arising from an APD's intrinsic gain mechanism \cite{apd_recent}.  In a typical APD consisting of a strongly reverse biased p-i-n junction, the applied electric field accelerates a photoinjected primary carrier until it impact ionizes, pulling another carrier across the semiconducting band-gap in order to create a multiplicative carrier gain. 

A key challenge in APDs is achieving high gain with low noise at longer wavelengths, where the material bandgaps start to approach the thermal energy. 
Over the years, the design and material engineering of APDs have grown highly sophisticated.  III-V digital alloy APDs have been reported to show low excess noise,  (multiplicative enhancement of shot noise) as well as high gain-bandwidth product, operating in the
short-wave infrared (SWIR) spectrum \cite{InAlAs_expt,AlInAsSb_expt,AlAsSb_expt}. These digital APDs consist of short-period superlattices with rich quantum mechanical properties such as the presence of minigaps, tunnel barriers and split-off valence bands, which play a strategic role in noise suppression by making the transport more unipolar, minimizing secondary ionizations \cite{apd_inequality, apd_strain}. Understanding and optimizing APDs need detailed simulation and design tools combining materials physics with carrier transport, all the way to physics based compact models \cite{ahmed_compact}. 

Impact ionization in bulk semiconductors has traditionally been simulated using ensemble Monte Carlo techniques. Electrons are treated as classical, Newtonian particles and their behavior modeled using semi-classical transport equations. Although these calculations use quantum ingredients like bandstructure, the transport is still classical and thus unsuited for explicit quantum effects like tunneling, or topological properties like spin-momentum locking. Furthermore, the carrier ionization rate is typically calculated using the Keldysh equation that incorporates the ionization threshold energy as a parameter \cite{zheng2020full} and the scattering probability itself as an empirical power law. For a homojunction APD with two parabolic bands, scalar effective masses $m_c$ and $m_v$ and a uniform bandgap $E_G$, the threshold energy can be estimated analytically using energy-momentum conservation laws \cite{ridley2013quantum}. However, the bandstructures of heterojunctions and digital alloy superlattices consist of a spaghetti of near-degenerate, non-parabolic and highly anisotropic energy bands, a proper treatment of which will necessitate energy and field-dependent effective mass tensors. In fact, in the absence of translational symmetry along the transport direction, these complexities strongly argue for a real-space, rather than a k-space treatment of transport. Unsurprisingly, Monte Carlo treatments reliant on constant masses tend to oversimplify the underlying chemistry, and need phenomenological quantum corrections to account for tunneling. A `first principles', predictive model that accounts for the complex materials chemistry, electrostatics and charge dynamics directly in real-space could be highly valuable in this regard.

In electronic device modeling, a fully quantum kinetic approach based on Non-Equilibrium Green's Functions (NEGF) \cite{datta_quantum_2005,datta2000nanoscale,Ghoshbook} has now been mainstream for decades. NEGF directly calculates the ensemble average of the non-equilibrium quantum mechanical electron charge density and current distributions, and is related to a histogram of single shot Monte Carlo results much the same way classical drift-diffusion equation relates to stochastic Newton's law (i.e., L\'{a}ngevin equation). In other words, it directly calculates the  quantum distribution functions, thermally averaged over locally equilibrium contact states.

The real strength of the NEGF approach is its matrix based formulation of Schr\"odinger equation with open, non-equilibrium boundary conditions at its bias-separated contacts. As a result, NEGF can directly incorporate a real-space Hamiltonian matrix that can account for sophisticated chemistry using either fully predictive `first principles' Density Functional Theory (DFT), or experimentally calibrated phenomenological tight binding (TB) approaches. A lot of commercial and open-source simulators have been based on TB-NEGF or DFT-NEGF - the former commonly used for device simulators {(e.g.~ Synopsys TCAD \cite{wu2018introduction}, NanoHUB \cite{klimeck2008nanohub})}, while the latter for molecular and nanoscale channel materials (e.g.~SIESTA \cite{soler2002siesta}, VASP \cite{hafner1997vienna}, Smeagol \cite{ferrer2014gollum}, Wien2K \cite{blaha2020wien2k}). When it comes to electron-electron scattering however, treatments of quantum transport lie at two extremes - weakly interacting electrons in mean-field (Poisson) approaches for electronic devices, or strongly correlated transport using multielecron master equations or configuration interaction theory for quantum dots \cite{muralidharan1,muralidharan2,muralidharan3,Ghoshbook}. To our knowledge, there has not been a matrix NEGF model that has been validated to capture strong Coulomb interactions underlying impact ionization in an APD, which is intrinsically a non-equilibrium, inelastic scattering dominated process requiring contraction of multiple Green's functions. {\it{In other words, in contrast to device models where inelastic scattering is at best a perturbative correction, APDs rely on impact ionization as a zeroth-order effect that needs a proper treatment.}} 

Quantum kinetic treatments of impact ionization in the literature have been largely limited to empirical fitting functions, more detailed treatments appropriate for isolated parabolic bands \cite{madureira2001impact,redmer2000field,Jacobini}, to small molecules \cite{Thygesen_GW_prb}, and by contracting Green's function products into an overall GW kernel \cite{Luisier_GW_iedm}. Generalizing it to matrix NEGF techniques is challenging because of the multiple carriers involved, Simpler, inelastic scattering due to phonons are captured in NEGF within a self-consistent Born approximation, where the electron in-scattering rate is proportional to a single electron or hole correlation function $G^{n,p}(E)$ \cite{datta_quantum_2005} - in effect, a spatially and energy resolved electron/hole density matrix. The traditional electron charge density is given by $n(x) = \int dE[G^{n}(E)]_{x,x}/2\pi$, analogously for holes. However, impact ionization involves collision between multiple carriers, which means that in-scattering here will involve multiple Green's functions, which to our knowledge has never been attempted so far.  

In this paper we develop a matrix NEGF description of impact ionization in APDs. We first introduce the way to include scattering self-energies within the NEGF framework. We next describe the self-energy for impact ionization in a minimal model of four energy levels corresponding to the four states involved before and after the scattering event. Finally, we extend the methodology to a model APD structure described with a Hamiltonian. The Hamiltonian we use corresponds to a simple one-dimensional chain of cross-linked dimers with tunable hopping parameters. We dope the chain to construct a p-i-n junction with photo-excitation (which we argue flips the polarities), and show that under a large reverse bias, our NEGF extracted charge and current distribution show impact ionization - namely, carrier multiplication at precise `dead-spaces' corresponding to energy-momentum conservation, an exponentially rising current and a gain by a factor of two at a single impact ionization site, if we suppress secondary ionization events. 

The advantage of this NEGF treatment is we can plot and visualize the spatial electron and hole dynamics. Furthermore, once the NEGF approach has been calibrated for our 1-D dimer Hamiltonian, it remains mainly a numerical exercise to generalize it to a more complicated Hamiltonian (e.g. conventional sp$^3$s$^*$d$^5$ approaches, Extended H\"uckel theory, or more accurate Environment Dependent Tight Binding), to add phonon scattering using conventional self-consistent Born approximation, and to extend to a fully 3-D structure by Fourier transforming the hopping parameters in the direction perpendicular to transport. We thus lay down the groundwork for developing a quantum transport model for multi dimensional nanoscale devices/materials that incorporates impact ionization.

\section{Non-Equilibrium Green's Function Method for inelastic transport} \label{NEGF}
\begin{figure}[h!]
\centering
\includegraphics[width=0.45\textwidth]{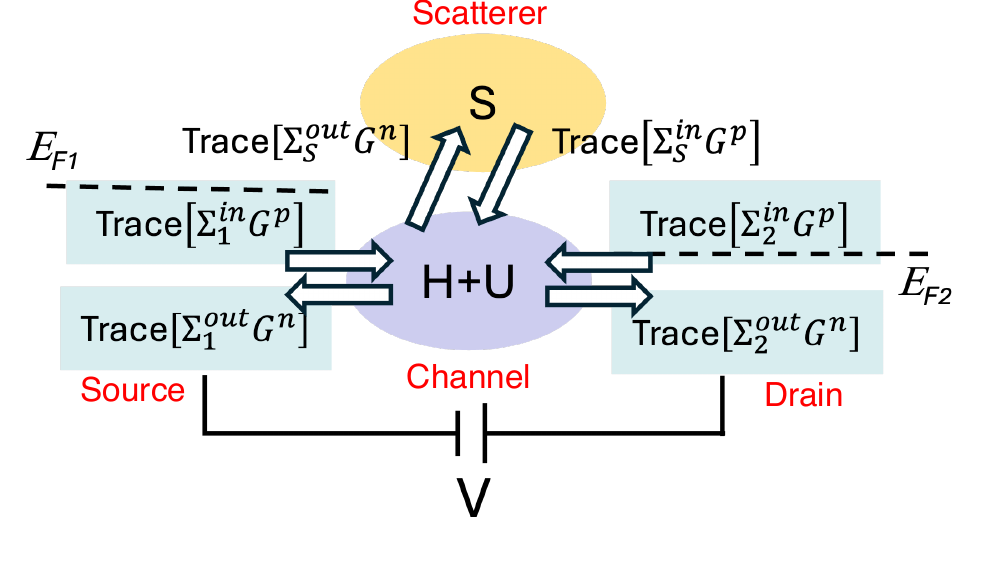}
\caption[Inflow and outflow in incoherent NEGF transport. ]{Inflow and outflow in non-coherent NEGF transport. At each terminal $\alpha$, influx is given by an electron in-scattering function $\Sigma^{in}_\alpha$ that feeds contact electrons into hole states $G^p$ in the channel, minus an out-scattering function $\Sigma^{out}_\alpha$ that siphons out channel electron states $G^n$. In-scattering $\Sigma^{in}_\alpha$ is typically given by a broadening matrix $\Gamma_\alpha$ that sets the escape rate into that terminal, times an occupation probability $f_\alpha$, while out-scattering is given by $\Gamma_\alpha(1-f_\alpha)$. For metallurgical contact terminals, $f_{1,2}$ are Fermi-Dirac distributions set by their local, voltage-separated electrochemical quasi-Fermi energies $E_{F1,2}$, while for scattering sites `s', the distribution $f_s$ is unknown, and we need a microscopic model for scattering to directly calculate $\Sigma^{in,out}_s$.}\label{fig:negf_nch}
\end{figure}
In this section, we explain the NEGF approach, and how to include self-energy matrices for various scattering events. We specifically write out the self-energy for phonon scattering, as has been traditional in the literature. We introduce the self-energy for impact ionization in the following section for current gain - which will be our main contribution to the literature. 

\subsection{The basic transport equations}\label{current_eqns}
In matrix based NEGF formalism (Fig.~\ref{fig:negf_nch}), the central channel is represented by a real space Hamiltonian $H$ that incorporates the material bandstructure, and a diagonal potential matrix $U$ that captures the electrostatic potential variation across the channel. 
Electron scattering happens at real metallurgical source-drain contacts `1,2' as well incoherent scattering sites `S' that act as virtual contacts, as shown in Fig. \ref{fig:negf_nch}.  The flow of electrons in and out of the contacts is captured by energy-dependent, self-energy matrices $\Sigma_{\alpha}(E)$,  that provide open (absorbing) boundary conditions at terminals $\alpha = 1,2, S$. These matrices are non-Hermitian, meaning their eigenvalues are complex numbers, their imaginary parts representing escape rates into the corresponding contacts. We will discuss their specific forms shortly.

When we project out the contact states, the resulting time-independent Schr\"odinger equation for the channel electrons naturally develops open boundary conditions with an inflow of electronic states $S_\alpha$ from the contacts, and an outflow given by $\Sigma_\alpha$ into the contacts
\begin{equation}
\Bigl[EI - H - U \underbrace{- \Sigma_1 - \Sigma_2 - \Sigma_S}_{\rm{Outflow}}\Bigr]\Psi = \underbrace{S_1 + S_2 + S_S}_{\rm{Inflow}}
\label{obse}
\end{equation}
The resulting open boundary Sch\"odinger equation is an inhomogeneous equation with a non-zero source on the right side. Its formal solution $\Psi$ is obtained by extracting the retarded Green's function $G$ that solves the equation for an impulse response (delta-function source)
\begin{eqnarray}
&&\Psi = G\Bigl(S_1+S_2 + S_s\Bigr)\nonumber\\
{\rm{where~}}&&\Bigl[EI - H - U -  \Sigma_1 - \Sigma_2 - \Sigma_s\Bigr]G = I \nonumber\\
\Longrightarrow && ~G = \Bigl[EI - H - U -  \Sigma_1 - \Sigma_2 - \Sigma_s\Bigr]^{\displaystyle -1}
\label{gfeqns}
\end{eqnarray}
Since the self-energies `open up' the system, their antiHermitian part gives us the broadening matrix related to escape rate
\begin{equation}
    \Gamma_\alpha = i(\Sigma_\alpha -\Sigma^\dagger_\alpha)
\end{equation}
while the antiHermitian part of the 
Green's function, the spectral function $A$, captures the local density of states $D(x,E)$ set along its diagonals
\begin{equation}
A = i(G-G^\dagger), ~~~ D(x,E) = [A(E)]_{x,x}/2\pi
\label{spectral}
\end{equation} 

Let us now move from static to dynamic properties of the channel electrons. The inflow and outflow processes are described by the additional in-scattering and out-scattering matrices $\Sigma_{\alpha}^{in}$ and $\Sigma_{\alpha}^{out}$ (Fig.~\ref{fig:negf_nch}), distinct from the retarded Green's functions $\Sigma_\alpha$ relevant for static properties. In Eq.~\ref{obse}, we assume that the source wavefunctions $S_{1,2}$ in separate contacts are uncorrelated thermodynamic variables and the contacts are set at local equilibrium, so that the bilinear thermal averages $\langle \ldots \rangle$ are set by the respective Fermi-Dirac distributions.$f_{1,2}(E) = 1/\left[1+e^{(E-E_{F1,2})/k_BT}\right]$, with local quasi-Fermi energies $E_{F1,2}$, whose difference defines the non-equilibrium boundary conditions. 
\begin{eqnarray}
\langle S_\alpha S^\dagger_\beta \rangle &=& \delta_{\alpha\beta}\Sigma_\alpha^{in}, ~~\alpha = 1, 2, s\nonumber\\
\Sigma^{in}_{1,2} &=& \Gamma_{1,2}(E)f_{1,2}(E)
\label{contuncorr}
\end{eqnarray}
Eqs. \ref{gfeqns},~\ref{contuncorr} then allow us to calculate the corresponding electron correlation function $G^n$, whose diagonal terms represent the space and energy resolved electron distribution
\begin{eqnarray}
G^n &=& \langle \Psi\Psi^\dagger\rangle = G\sum_{\alpha\beta}\underbrace{\langle S_\alpha S^\dagger_\beta\rangle}_{\displaystyle \delta_{\alpha\beta}\Sigma^{in}_\alpha} G^\dagger\nonumber = G\Sigma^{in} G^\dagger\\
\Sigma^{in} &=& \sum_\alpha \Sigma^{in}_\alpha \nonumber\\
n(x,E) &=& [G^n(E)]_{x,x}/2\pi, ~~~ n(x) = \int dEn(x,E) 
\label{eqkel}
\end{eqnarray}
and a corresponding hole correlation function $G^p$. 

From the electron and hole correlation functions and the various in and out-scattering functions, we can calculate the current at any contact using the Meir-Wingreen formula \cite{Ghoshbook}
\begin{equation}
    I_\alpha = \frac{q}{h} \int dE Tr \left[\Sigma_\alpha^{in}(E)G^p(E)-\Sigma_\alpha^{out}(E) G^n(E)\right] 
    \label{mw}
\end{equation}
which states that the incoming current involves in-scattering $\Sigma^{in}_\alpha$ into empty (hole) states $G^p$, while the outgoing current involves out-scattering $\Sigma^{out}_\alpha$ from filled electronic states $G^n$ (Fig.~1).

\subsection{Calculating the scattering matrices $\Sigma_\alpha$, $\Sigma^{in,out}_\alpha$}
The electron and hole correlation functions are obtained from the Keldysh equation (first part of Eq.~\ref{eqkel}), $G^{n,p} = G\Sigma^{in,out}G^\dagger$, summing over the individual in/out-scattering matrices $\Sigma^{in,out} = \sum_\alpha\Sigma^{in,out}_\alpha$. The scattering matrices however need a model based treatment dependent on their microscopic origin.

\textcolor{black}{The source/drain contacts come with their own Hamiltonians, which decompose naturally into a block tridiagonal form involving diagonal onsite and off-diagonal hopping matrices between their unit cells, from which the self-energies $\Sigma_{1,2}$ and their antiHermitian broadening matrices $\Gamma_{1,2}$ can be calculated using recursion, exploiting the semi-periodic nature of each contact \cite{datta2000nanoscale,Ghoshbook,Ghoshbook2}. The resulting broadening matrix follows Fermi's Golden Rule, involving the hopping matrices between channel and contacts, and the spectral function of the surface states.}
For source-drain contacts with inflow terms in Eqs.~\ref{contuncorr},\ref{eqkel}, the in and out scattering terms are set by the broadening matrices, as argued above \begin{eqnarray}
\Sigma^{in}_{1,2} &=& \Gamma_{1,2}f_{1,2}\nonumber\\
\Sigma^{out}_{1,2} &=& \Gamma_{1,2}(1-f_{1,2})
\end{eqnarray} where $f_\alpha = f(E-E_{F\alpha})$ are the local Fermi-Dirac distributions of the electrons in the contacts, assumed to be reservoirs in local equilibrium. 
However, since there is no externally imposed Fermi function  describing the `virtual' scattering terminal, there is no simple connection between $\Sigma_S^{in,out}$ and $\Gamma_S$, nor default expressions for either. We need a microscopic model for these scattering processes. 

The self-energy for incoherent scattering such as from acoustic phonons is usually captured within the self-consistent Born approximation. The in and out scattering functions for the virtual terminal at a particular energy can be generally written as \cite{datta2018lessons,Ghoshbook}: 
\begin{eqnarray}
 \Sigma_S^{in}(E) &=& D \otimes G^n(E) \nonumber\\
 \Sigma_S^{out}(E) &=& D \otimes  G^p(E)
 \label{scba1}
\end{eqnarray}
where, the components of the deformation potential $D_{ij} =\langle U_i U_j^* \rangle$ represent the ensemble average of the correlation between the random interaction potentials at the points $i$ and $j$. The $\otimes$ sign means an element by element matrix multiplication. 
$D$ is actually a fourth rank tensor, $\Sigma^{in}_{ij} = \sum_{kl} D_{ijkl}G^n_{kl}$, accounting for non-locality of the interaction potential $D_{ijkl} = \langle U_{ik}U^*_{jl}\rangle$ contracted over the indices of the scattering center and averaged over its thermal distribution, the scatterers assumed to be in local equilibrium with an underlying Fermi-Dirac or Bose-Einstein distribution \cite{Ghoshbook}. When the device size is longer than the extent of non-locality, we can replace the $U$ (electron-phonon coupling, or screened Coulomb potential) by a local approximation $U_{ij} \approx U_i\delta_{ij}$, leading to the expression above.

If we include inelastic scattering by optical phonons of frequency $\omega$, then the equation modifies to account for both phonon emission and absorption as
\begin{eqnarray}
 \Sigma_S^{in}(E) &=& D \otimes \Bigl[G^n(E+\hbar\omega)(N_\omega + 1) + G^n(E-\hbar\omega)N_\omega\Bigr] \nonumber\\ 
 \Sigma_S^{out}(E) &=& D \otimes  \Bigl[G^p(E-\hbar\omega)(N_\omega + 1) + G^p(E+\hbar\omega)N_\omega\Bigr] \nonumber\\
 &&
\label{scba2}
\end{eqnarray}
where $N_\omega = \Bigl[e^{\hbar\omega/kT}-1\Bigr]^{-1}$ is the Bose-Einstein distribution that sets the equilibrium phonon emission probability $N_\omega + 1$ and absorption probability $N_\omega$, with the extra unity term (spontaneous emission) enforcing a Boltzmann ratio between absorption and emission, $N_\omega/(N_\omega + 1) = \exp{[-\hbar\omega/kT]}$. For a distribution of phonons, the above equation needs to be summed over the phonon density of states, $\int d\omega D_{ph}(\omega)$.

Notice that the very structure of the self-consistent Born approximation (Eq.~\ref{scba1},~\ref{scba2}) guarantees that the current drawn by any scattering terminal $I_S(E)$ (Eq.~\ref{mw} with $\alpha = S$) will involve  a trace of $G^nG^p - G^pG^n$ in Eq.~\ref{mw} adding up to zero, so that the phonons do not draw any net current, $I_S = 0$, and in consequence the source and drain currents are equal and opposite, $I_1 = -I_2$, as expected from Kirchhoff's law. %\textcolor{blue}{In device modeling, it is common to invoke a so-called B\"uttiker probe approximation that constructs the scattering processes in a way to enforce this vanishing of the scattering current. In our approach, however, we will see that the current conservation is guaranteed by the structure of the matrices.} 

From the model scattering functions, we can also calculate the broadening matrix due to scattering, and the corresponding scattering self-energy that needs to obey Kramers-Kr\"onig equation to conserve spectral weight in energy and causality in time
\begin{eqnarray}
\Gamma_S &=& \Sigma^{in}_S + \Sigma^{out}_S \nonumber\\
\Sigma_S &=& {\cal{H}}(\Gamma_S) - i\Gamma_S/2
\end{eqnarray}
where ${\cal{H}}$ denotes a Hilbert transform.

The entire coupled set $G, G^{n,p}, \Sigma_\alpha, \Sigma^{in,out}_\alpha$ depend on each other and need to be calculated self-consistently.

\textcolor{black}{While phonon scattering has been routinely included in NEGF calculations for current in nanotransistors, and is, in fact, consequential in APDs (e.g. relaxing the hot electrons and postponing the onset of ionization), in this paper, we will focus exclusively on electron impact ionization, clearly the most involved of the lot, and ignore other inelastic processes.}

\subsection{Self-energy for impact ionization}
\begin{figure}[b]
\centering
\includegraphics[width=0.25\textwidth]{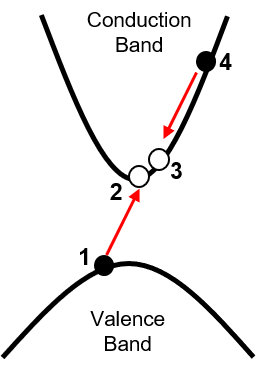}
\caption[Schematic of electron impact ionization.]{Schematic of electron impact ionization. The collision between a hot electron 4 and an electron 1 in the valence band scatters them to two empty energies 2, 3, ending with two cold electrons near the band-bottom, and holes where 1 and 4 sat.}\label{fig:electron_ionize}
\end{figure}

We now write down the scattering matrices corresponding to impact ionization. Since the transport is bipolar, we define $\Sigma_S^{in}$ to describe the inscattering of majority carriers (electrons for n-doped, holes for p-doped semiconductors), and $\Sigma_S^{out}$ for the outflow of majority carriers or inflow of minority carriers. We use Fig.~\ref{fig:electron_ionize} to guide our expression. 

For electron impact ionization, an electron injected in the conduction band gains kinetic energy from the applied electric field at reverse bias. At one point (Fig. \ref{fig:electron_ionize}) the scattering self-energies compel the `hot' electron to drop down closer to the conduction band-edge and transfer its excess kinetic energy to an electron in the valence band, lifting it across the bandgap into the conduction band, thereby multiplying the electron current in the conduction band and adding a hole current in the valence band. The process involves four energy states - three in the conduction band and one in the valence band. In the literature, impact ionization scattering terms are related to the carrier concentrations as $n^2p$ or $p^2 n$, with a fourth term being the majority electron concentration in valence band or hole concentration in conduction band, essentially a constant. For the matrix based NEGF theory, these relationships will be described in terms of the electron concentration $G^n_{c,v} = G^n\otimes [\Theta]$ and hole concentration $G^p_{c,v} = G^p\otimes [\Theta]$ in the corresponding band. 

\textcolor{black}{Keeping in mind that the scattering terms involving four energy states (before and after) separate into in and out scattering functions $\Sigma^{in,out}$ times the corresponding occupancy $G^{p,n}$ (Eq.~\ref{mw}), we expect that each scattering term itself will involve three terms. For impact ionization between two bands $(b,b')$, we can, by inspection, write the in-scattering and out-scattering self-energies in compact notation as}
\begin{eqnarray} 
    \Sigma_{S,b}^{in} &=&  D_0\otimes [G_b^{min}*G_b^{maj}G_{b^\prime}^{maj}] \label{eq:sigma_in} \nonumber \\
     \Sigma_{S,b}^{out} &=& D_0\otimes  [G_{b^\prime}^{maj}*G_b^{min}G_{b}^{min}] \label{eq:sigma_out}
    \label{eqii}
\end{eqnarray}
where the  operation $[A*BC](E)$ is defined as:
\begin{eqnarray}
    &&[A*BC](E) = 
    \iiint  dE^\prime dE^{\prime\prime} dE^{\prime\prime\prime} \nonumber\\
&&A(E^\prime)B(E^{\prime\prime})C(E^{\prime\prime\prime})\delta(E+E^\prime-E^{\prime\prime}-E^{\prime\prime\prime})\nonumber
\end{eqnarray}
with the symbol $*$ included to keep track of the signs of the energy terms. 
From the entire Green's functions across all bands, an individual band can be picked out by invoking a $\Theta$ function that is basically a diagonal matrix filtering out the relevant energies relative to the position-dependent band-edges. 
\begin{equation}
G_b = G \otimes \Theta(E\in E_b)
\end{equation}
where $\Theta$ is a diagonal matrix with components 0 or 1 depending on whether the energy $E$ belongs to the local band at that energy, e.g., above the position dependent band-edge $E_c(x)$ for conduction band or below $E_v(x)$ for the valence band.
For a Hamiltonian with multiple orbitals on each atom, we use the same $x_i$ for all orbitals at one atomic coordinate, so that for $N$ spatial grid points with $n$ orbitals per grid point, all the matrices above are of size $Nn \times Nn$.

\textcolor{black}{Note that keeping the  band indices distinct, $b\neq b^\prime$, explicitly discounts intraband processes. In practical calculations however, it may be expedient to sum over all bands including intraband scattering, since the latter does not make a difference to the computed impact ionization current.} However for amorphous structures with enormous band-mixing, intraband processes that are band unrestricted will start to matter, degrading the APD performance (the dark current and thermal noise will be unacceptably large). Our focus is on practical APD heterostructures with a clear band-gap, where the concept of `local bands' invoked above, with the $\Theta$ function, is justified. In order to include intraband processes, we will need to drop the band indices in Eq.~\ref{eqii}, but doing so for materials with band-gaps would necessitate special attention to the energy-dependence of the $D_0$ matrices to distinguish intra and interband processes.

The goal of the $[\Theta]$ matrix term is to limit the $G^{n,p}$ correlation functions to only the bands where carriers form a minority, ie, electrons $n$ in the conduction band and holes $p$ in the valence band. Note that the band-gap region $E_v(x_i) < E < E_c(x_i)$ yields zero contribution, which is expected because as we will see later, electrons will be photo-injected into the conduction band and removed after impact ionization from the valence band. In other words, each energy $E$ lies within only one band and only one of the Heaviside terms is activated at a time. 

As before, the scattering current can be written by summing individual band contributions
\begin{equation}
    I^{maj}_{S} = \int dE Tr[\Sigma^{in}_{S}G^{min} - \Sigma^{out}_{S}G^{maj}]
    \label{iii}
\end{equation} 
Notice that \textcolor{black}{when applied across two bands}, the two end terms \textcolor{black}{in Eq.~\ref{eqii}} are $G_b^{min}$, $G^{maj}_{b^\prime}$, \textcolor{black}{so that in the final product of four terms, we get two terms for each carrier (maj, min), but three terms from one band $b$ or $b^\prime$, and one from the other (Fig.~\ref{fig:electron_ionize}). 
It is easy to see that when included with
$\Sigma^{in,out}_{S,b}$ from Eq.~\ref{eqii}}, the terms in the middle look like $G^{maj}_bG^{min}_b-G^{min}_bG^{maj}_b$, which upon tracing with the sandwich terms give $I_S = 0$. The impact ionization will then show up as an exponential increase in carrier concentration at the drain end $I_2 = -I_1$ compared to ballistic current.  
% \begin{equation}
%     [A*BC](E) = D_0 \otimes \iiint dE' dE'' dE''' A(E')B(E'')C(E''') \delta(E+E'-E''-E''')
% \end{equation}\\
% Looking at the two end terms of the $\Sigma_S$ that remain the same, we see clearly that the term in the middle looks like $G_b^{maj}G_b^{min}-G_b^{min}G_b^{maj}$, which upon tracing with the sandwich terms gives zero. 

\section{Impact Ionization in a Four-level System}

\begin{figure}[b]
\centering
\includegraphics[width=0.48\textwidth]{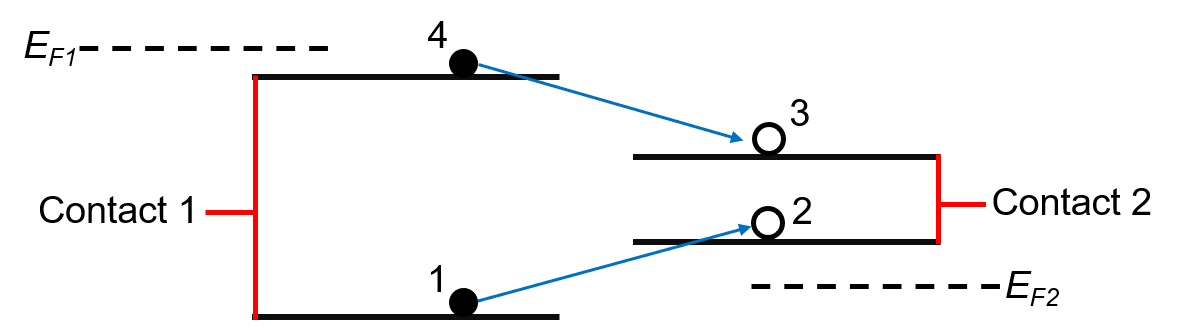}
\caption[Schematic of four level system under impact ionization.]{Schematic of a four level system under impact ionization. The energy levels are chosen so as to satisy energy conservation, $\epsilon_4-\epsilon_3=\epsilon_2-\epsilon_1$.}\label{fig:four_level_schem}
\end{figure}

Let us unpack the process of impact ionization with a minimal model, namely a four-level system shown in Fig. \ref{fig:four_level_schem}. We designed the system such that the initial states are connected to contact 1 and the final states are connected to contact 2. The system consists of four energy levels whose onsite energies are denoted by $\epsilon_1,\ldots,\epsilon_4$. There is no direct coupling between the different energy levels. This ensures there is no current flowing through the system under ballistic conditions, {even under strong voltage bias}. The four level Hamiltonian can be written as:

\begin{equation}
H=\left(\begin{array}{cccc}
\epsilon_1 & 0 & 0 & 0\\
0 & \epsilon_2 & 0 & 0\\
0 & 0 & \epsilon_3 & 0\\
0 & 0 & 0 & \epsilon_4\\
\end{array}\right)    
\end{equation}

Current flows through this system only when the conditions of electron impact ionization are satisfied. The quasi-Fermi levels/electrochemical potentials at the two contacts, $E_{F1}$ and $E_{F2}$, need to be set such that electrons are injected into the device from the left contact and are extracted from the right contact. In the system, $E_{F1}=E_{F0}+V/2$ and $E_{F2}=E_{F0}-V/2$, where $V$ is the applied voltage across the terminals and $E_{F0}$ is the equilibrium Fermi level of the system. {This convention for the movement for the quasiFermi energies assumes equal capacitive coupling to the two contacts, so that the channel states shift on average by half the applied bias \cite{ghoshpot}} For impact ionization to happen, $E_{F1}$ must be above the levels $1$ and $4$, and $E_{F2}$ should be below $2$ and $3$. Electrons are then injected into the low energy state $1$ and the high energy state $4$. These electrons move to the empty states at the energy levels $2$ and $3$, respectively, due to the impact ionization process, and are swept away by the right contact. The carriers must satisfy energy conservation, \textit{i.e.}, $\epsilon_4-\epsilon_3=\epsilon_2-\epsilon_1$, so the energy levels are specifically chosen in this example to satisfy this energy conservation. 

The next step involves defining the in/out scattering functions $\Sigma_S^{in}$ and $\Sigma_S^{out}$ for impact ionization in  this system following the expressions introduced earlier (Eq.~\ref{eqii}). Electrons enter from the states $1$ and $4$ into states $2$ and $3$ when the above mentioned conditions are satisfied. Thus, we need outscattering functions for the states 1 and 4, and inscattering functions for the states 2 and 3. These functions can be expressed as: 
\begin{eqnarray}\label{sigma_frlvl} \nonumber
\Sigma_S^{out,1}(E) &=& D \otimes \int dE^{'''} dE^{''}dE^{'} G^p_2(E^{'''})G^p_3(E^{''})G^n_4(E^{'})\nonumber\\ 
&& \delta(E^{'''}-E-E^{'}+E^{''})\\ \nonumber
\Sigma_S^{out,4}(E) &=& D \otimes \int dE^{'''} dE^{''}dE^{'} G^p_2(E^{'''})G^p_3(E^{''})G^n_1(E^{'})\nonumber \\ 
&& \delta(E^{'''}-E^{'}-E+E^{''})\\ \nonumber
\Sigma_S^{in,2}(E) &=& D \otimes \int dE^{'''} dE^{''}dE^{'} G^n_1(E^{'''})G^n_4(E^{''})G^p_3(E^{'}) \nonumber \\
&& \delta(E-E^{'''}-E^{''}+E^{'})\\ \nonumber
\Sigma_S^{in,3}(E) &=& D \otimes \int dE^{'''} dE^{''}dE^{'} G^n_1(E^{'''})G^n_4(E^{''})G^p_2(E^{'})\nonumber \\ 
&& \delta(E^{''}-E-E^{'}+E^{'''})
\end{eqnarray}
where, $D$ is treated as a multiplicative constant for now. The indices 1 to 4 represent the four states in the system. Energy conservation is satisfied by the delta functions. The scattering terminal $I_{S}$ current (Eq.~\ref{mw}) can then be written as
\begin{eqnarray}\label{isc_frlvl}
    I_{S} &=&\int dETr\left[\Sigma_S^{in,2}(E)G^p_2(E)+\Sigma_S^{in,3}(E)G^p_3(E)\right.\nonumber\\
    && \left.-\Sigma_S^{out,1}(E)G^n_1(E)-\Sigma_S^{out,4}(E)G^n_4(E)\right]
\end{eqnarray}
which as expected vanishes, $I_{S}=0$. $\Sigma_S^{in}$ and $\Sigma_S^{out}$ matrices are then expressed using the equations: 
\begin{equation}
\Sigma_S^{in}(E)=\left(\begin{array}{cccc}
0 & 0 & 0 & 0\\
0 & \Sigma_S^{in,2}(E) & 0 & 0\\
0 & 0 & \Sigma_S^{in,3}(E) & 0\\
0 & 0 & 0 & 0\\
\end{array}\right)    
\end{equation}

\begin{equation}
\Sigma_S^{out}(E)=\left(\begin{array}{cccc}
\Sigma_S^{out,1}(E) & 0 & 0 & 0\\
0 & 0 & 0 & 0\\
0 & 0 & 0 & 0\\
0 & 0 & 0 & \Sigma_S^{out,4}(E)\\
\end{array}\right)    
\end{equation}

\begin{figure}[t]
\centering
\includegraphics[width=0.4\textwidth]{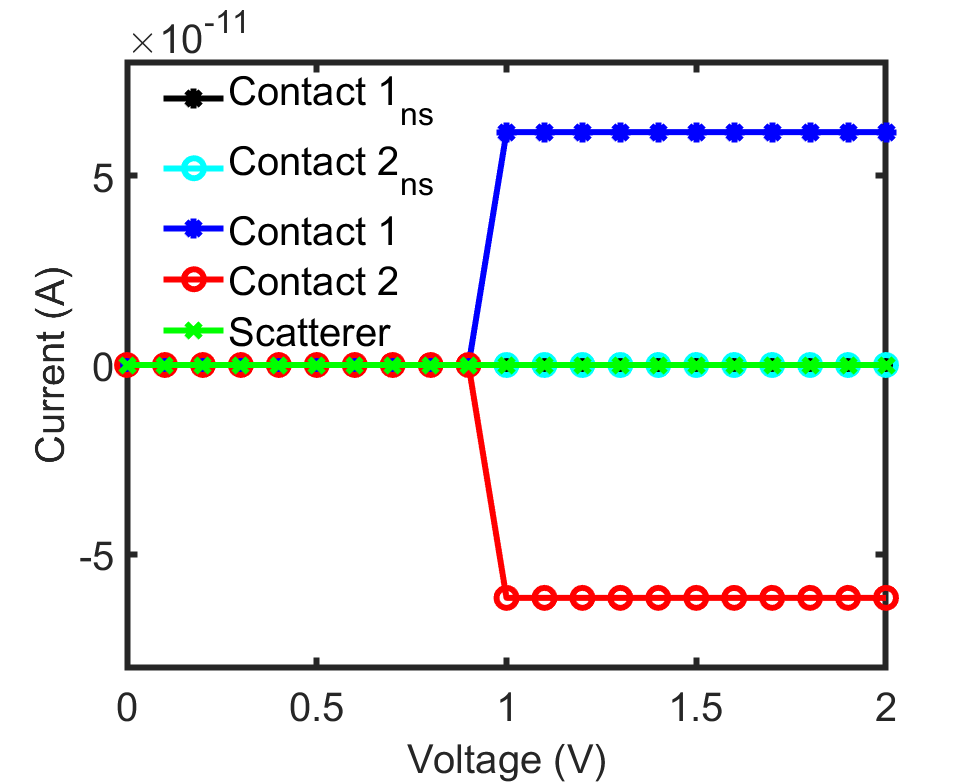}
\caption[Current vs. Voltage characteristics of a four-level system.]{Current vs. Voltage characteristics of a four-level system. The scattering current $I_S = 0$, as are the in-coming currents in contacts 1 and 2 in the absence of scattering. {At $\sim 0.8$ volt applied bias, we inject electrons into state 4 from contact 1 (Fig. ~\ref{fig:four_level_schem}), and remove them from state 2 using contact 2, and inelastic scattering `bridges the gaps' between the discrete energy levels by moving the filled valence electron in 1 and injected electron in 4 into states 2 and 3.} }\label{fig:four_level_IV}
\end{figure}

\begin{figure}[b]
\centering
\includegraphics[width=0.4\textwidth]{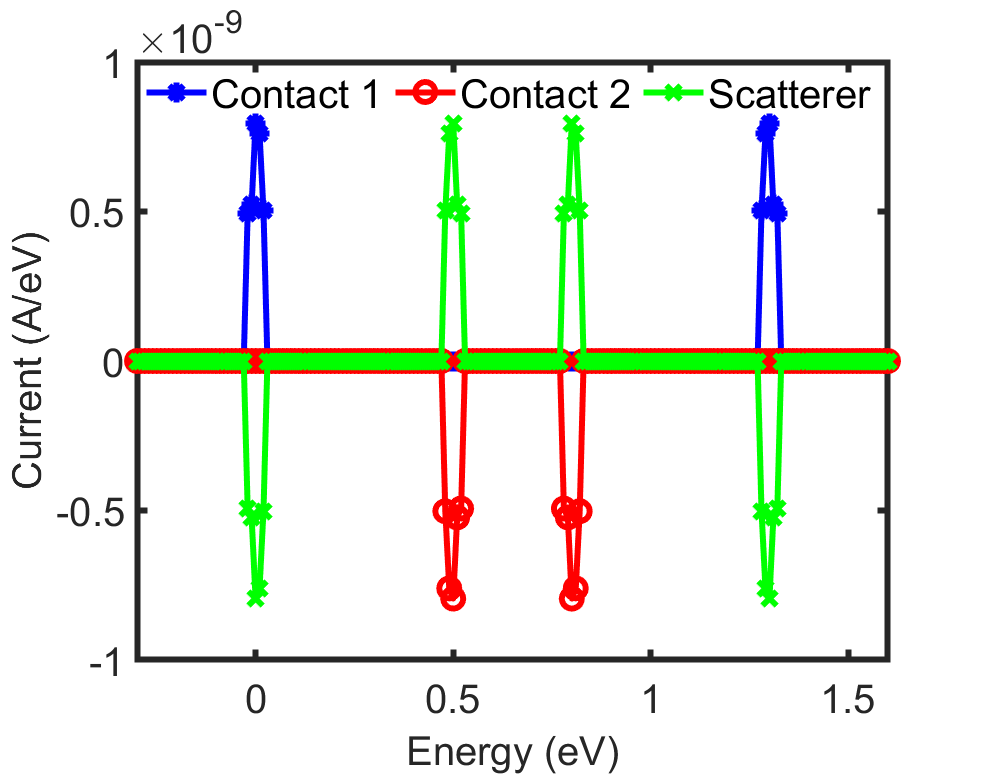}
\caption[Energy resolved current for a four-level system at $V=1.5V$.]{Energy resolved current for a four-level system at $V=1.5V$. The scattering currents allow energy redistribution of incoming contact 1 currents at energies $\epsilon_{1,4}$ to outgoing currents in contact 2 at energies $\epsilon_{2,3}$, while itself adding up to zero.}\label{fig:four_level_singlebias}
\end{figure}

Finally, the terminal currents of the four-level system are computed using the equations described in Section \ref{current_eqns}, namely Eq.~\ref{mw}. Fig. \ref{fig:four_level_IV} plots this terminal current. We can see that under ballistic, non-scattering (`ns') conditions, there is no current flowing through the system (Contact $1_{ns}$ and Contact $2_{ns}$ currents are zero). The terminal current including impact ionization shows a sharp jump at $V=1V$. At this voltage $E_{F1}$ is above state 4 and $E_{F2}$ is below state 2 and energy conservation is also satisfied, resulting in the jump. The $I_{S}=0$ condition is also satisfied in this plot. For our simulations, we set $\epsilon_1=0~eV$, $\epsilon_2=0.5~eV$, $\epsilon_3=0.8~eV$, $\epsilon_4=1.3~eV$, satisfying energy conservation, $E_{F0}=0.86$ eV, $D=10$, and temperature $T=3$ K. The energy resolved current is plotted in Fig. \ref{fig:four_level_singlebias}. We can see that the left contact (blue) injects electrons in the states 1 and 4. The virtual contact (green) takes out these electrons and reinserts them into states 2 and 3, {bridging the energy gap through impact ionization}, and the electrons from states 2 and 3 are then carried away by the right contact (red). {Compared to the ballistic current which was zero, the terminal current in presence of impact ionization has now dramatically increased}.

Having calibrated our NEGF model to satisfy a minimal four-level system, we will now extend the model to a 1D semiconductor. 
\section{Impact ionization in a 1D semiconductor}\label{II1D}

\begin{figure}[htbp]
\centering
\includegraphics[width=0.45\textwidth]{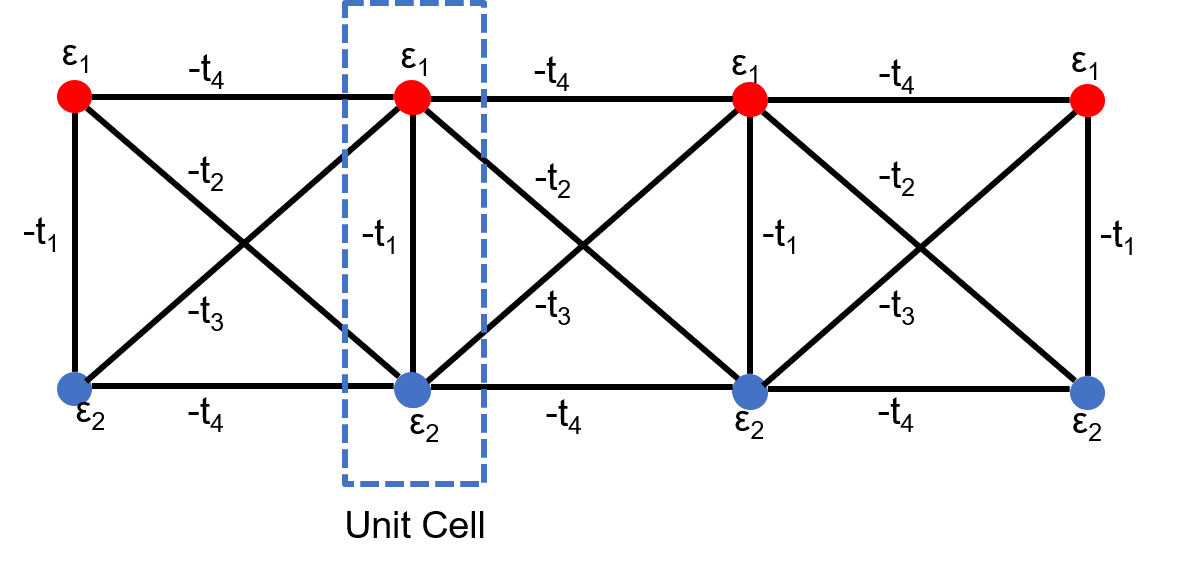}
\caption[One dimensional cross linked dimer chain structure with parabolic bands.]{One dimensional cross linked dimer chain structure with parabolic bands.}\label{fig:cross_dimer}
\end{figure}

Bulk semiconductors have energy bands instead of discrete energy levels. Here, we extend the matrix based NEGF theory for impact ionization to such a material. We study a one dimensional dimer chain with two energy bands that are near parabolic (energy-independent effective masses over a fairly large band-width). Since the threshold energy for parabolic bands can easily be calculated with an analytical expression, $E_{TH} = E_G(1+2\mu)/(1+\mu)$, with $\mu = m^*_c/m^*_v$ for primary electron injection (inverse for hole), it is easy to test the validity of our model. The schematic of a one dimensional cross-linked dimer chain, is shown in Fig. \ref{fig:cross_dimer}, along with its unit cell.  The two onsite energies and four cross-linked parameters allow us to tune the effective mass ratios between the conduction and valence bands. 

The dimer chain provides a versatile model for direct band-gap semiconductors (Table~\ref{tab:material-params} and Fig.~\ref{fig:Impact_Ionization_all_III-V} later show the applicability to multiple binary APDs). The matrix nature of its Hamiltonian blocks prepares our numerical simulation for orbital based sp$^3$s$^*$ type atomistic models in future. As it stands however, it is a compromise between fully atomistic bandstructures with oversimplified scattering (constant $D_0$), vs oversimplified bandstructures (scalar effective mass) with involved Golden Rule type scattering terms.\\

\subsection{1-D Dimer chain Hamiltonian}

\begin{figure}[t]
\centering
\includegraphics[width=0.4\textwidth]{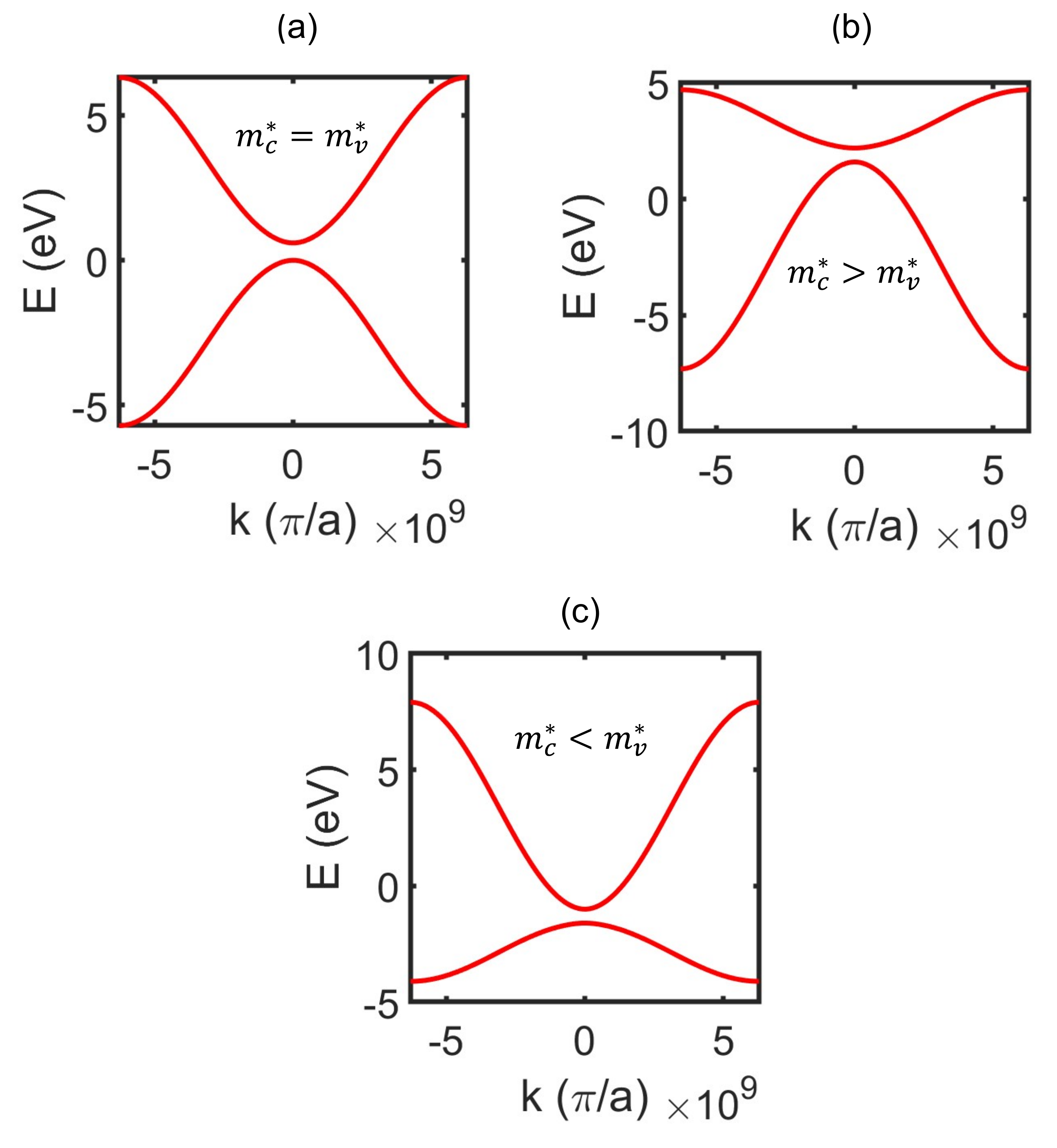}
\caption[Bandstructure for the dimer chain for (a) $m_c^*=m_v^*$ (b) $m_c^*>m_v^*$ and (c) $m_c^*<m_v^*$.]{Bandstructure for the dimer chain for (a) $m_c^*=m_v^*$ (b) $m_c^*>m_v^*$ and (c) $m_c^*<m_v^*$.}\label{fig:cross_dimer_Ek}
\end{figure}

The Hamiltonian for a 1D chain becomes 
\begin{equation}
H=\left(\begin{array}{cccccc}
\alpha & \beta & 0\\
\beta^{+} & \alpha & \beta & 0\\
0 & \beta^{+} & \alpha & \beta\\
 & 0 & \cdots & \cdots & \cdots\\
 &  &  &  & \cdots & \beta\\
 &  &  &  & \beta^{+} & \alpha
\end{array}\right)
\end{equation}
in a dimer basis set comprising the unit cell outlined in a blue dashed box (Fig.~\ref{fig:cross_dimer}), and the onsite (intra-cell) and hopping (inter-cell) Hamiltonian blocks separated by period $a$ are given by
\begin{equation} \nonumber
\alpha=\left(\begin{array}{cc}
\epsilon_{1} & -t_1\\
-t_1 & \epsilon_{2}
\end{array}\right),~~~~\beta=\left(\begin{array}{cc}
-t_4 & -t_2\\
-t_3 & -t_4
\end{array}\right).
\end{equation}
\\
The resulting dimer bandstructure, obtained by plotting eigenvalues of the Fourier transformed Hamiltonian, $H_k = \sum_n[H]_{mn}e^{ik(m-n)a}$ is shown in Fig. \ref{fig:cross_dimer_Ek}. We can create asymmetry between the conduction band and valence band effective masses by varying the different couplings in the Hamiltonian. For our simulations, we set $\epsilon_1=0.6$, $\epsilon_2=0$, $t_1=3$, $t_2=-1$ and $t_3=-2$ eV. We can then vary the value of $t_4$ to create the mass asymmetry. Fig. \ref{fig:cross_dimer_Ek}(a) shows the bandstructure for equal mass for which $t_4=0$. In Fig. \ref{fig:cross_dimer_Ek}(b) we set $t_4=-0.8$ which results in $m_c^*>m_v^*$, while $t_4=0.8$  causes $m_c^*<m_v^*$ as in Fig. \ref{fig:cross_dimer_Ek}(c).  The bandgap is the same for all three cases. 

\subsection{Band-diagram under photo-injection in APD}

\begin{figure}[b]
\centering
\includegraphics[width=0.45\textwidth]{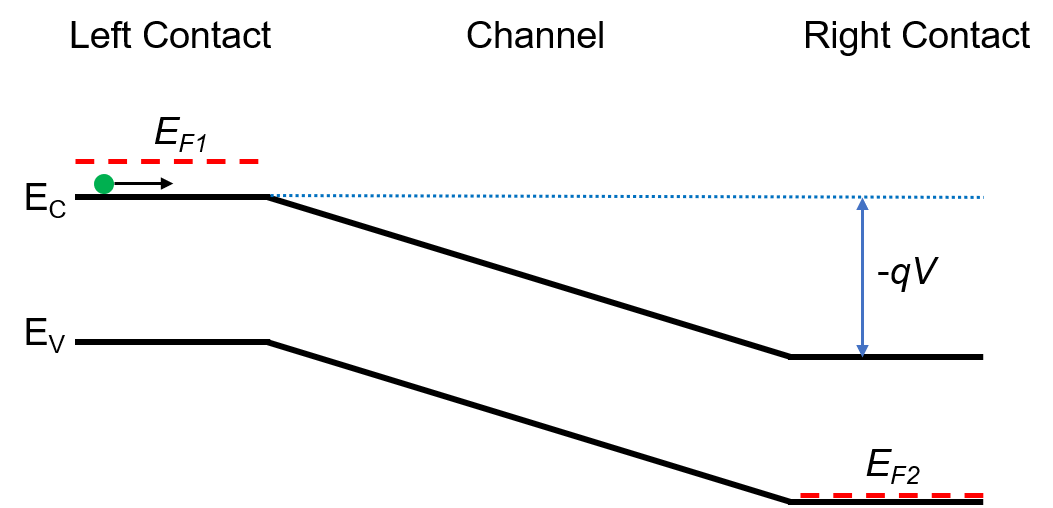}
\caption[Potential diagram of a 1D semiconductor device for studying impact ionization.]{Potential diagram of a 1D semiconductor device for studying impact ionization.}\label{fig:1D_device}
\end{figure}

A typical APD starts with a p-i-n structure under strong reverse bias, to create a large electric field to separate the charges. Many APDs come embedded within a separate absorber charge multiplier (SACM) configuration, where primary carriers are resonantly photo-excited across the bandgap in the separate absorption (SA) region outside the APD, and then injected into the active charge multiplication (CM) region where they are accelerated by the applied field for impact ionization. We will focus on just the CM region, but account for the injection from the SA region by placing our quasi-Fermi levels accordingly. For an n-type APD that we simulate, electrons will be injected from the source p-region into the conduction band, meaning the quasi-Fermi energy must be in the conduction band of the p region. The electrons are then pulled out after impact ionization so the quasi-Fermi energy in the drain must be near or below the conduction band-edge to extract all the charges. We place it near the valence band, although it could sit anywhere in the bandgap without a difference in current. 
In effect therefore, our band-diagram is a reverse biased p-i-n junction, but the Fermi level placements resemble an n-i-p structure (Fig.~\ref{fig:1D_device}). 

A simple NEGF model can capture the SA absorption physics in a p-region that promotes its quasi-Fermi level from valence band to conduction band. We need to consider a dimer model without any bias to represent the external bias-free separate absorption (SA) region outside the active reverse-biased carrier multiplication (CM) region, and place the SA quasi-Fermi energy in the valence band to calculate its correlation function $G^n = Af_0$, $A$ being its spectral function (Eq.~\ref{spectral}). $G^n$ at this point will have very little weight in the conduction band. We now apply photons of energy $\hbar\omega$ slightly above the band-gap, which creates a scattering event with in-scattering function $\Sigma^{in}_{sc}(E) = D_{ph}G^n(E-\hbar\omega)N_\omega$, Eq.~\ref{scba2}, and recompute the correlation function $G^n = G\Sigma^{in}_{sc}G^\dagger$, Eq.~\ref{eqkel}, which will now shift the charge distribution to the conduction band. Finally, we compute the effective Fermi-function $f_{eff}(E) = G^n(1,1)/A(1,1)$ at point 1, or any of the equivalent positions, and fit it to a Fermi-Dirac distribution to extract the quasi-Fermi energy, which by now will have visibly shifted to the conduction band. This calculation needs to be done self-consistently. The photon flux responsible for $N_\omega$ is obtained from the input power divided by the photon energy, while the parameter $D_{ph}$ can be obtained from electron-photon coupling in the SA region. {The end result is that the Fermi level placements will resemble an inverted, n-i-p junction under high forward bias, rather than a p-i-n junction, under high reverse bias (Fig.~\ref{fig:1D_device})}.

\newpage
\subsection{Vanishing of the scattering current}

Before we compute the inelastic current contribution at the terminals, we need to verify that the scattering current is zero. For electron impact ionization in a semiconductor, three of the energy states are in the conduction band - the high-energy electron and the two empty states into which electrons flow after ionization (Fig.~\ref{fig:electron_ionize}). The remaining low energy state is in the valence band. We need to distinguish between these states when we extend the model of the four-level system to that with conduction and valence bands. This is done by setting limits to the integrals in the equation of the virtual scattering terminal current. For a semiconductor with energy bands, we can write $I_{S}$ from Eq.~\ref{iii} as
\begin{widetext} 
\centering
\begin{eqnarray}\label{Isc_2band}
I_{S} &=&  \int dE_2Tr \Sigma_S^{in}(E_2)G^p_c(E_2) + \int dE_3 Tr\Sigma_S^{in}(E_3)G^p_c(E_3)-  \int dE_4 Tr\Sigma_S^{out}(E_4)G^n_c(E_4)- \int dE_1 Tr\Sigma^{out}_S(E_1)G^n_v(E_1) \nonumber\\
&&
\end{eqnarray}
\end{widetext}
where
\begin{widetext}
\centering
\begin{eqnarray}
\label{sig_2band}
% {\rm{where~~}}
\Sigma^{in}_S(E_{2,3} > E_{cR}) &= D \otimes \int dE_{3,2} G^p_c(E_{3,2}) \int dE_4 G^n_c(E_4) \int dE_1 G^n_v(E_1)  \delta(E_4-E_3-E_2+E_1)\\  \nonumber  
     \Sigma^{out}_S(E_{4} > E_{cR}) &= D \otimes \int dE_2 G^p_c(E_2) \int dE_3 G^p_c(E_3) \int dE_{1} G^n_v(E_{1}) \delta(E_4-E_3-E_2+E_1)\\ \nonumber
      \Sigma^{out}_S(E_{1} < E_{vL}) &= D \otimes \int dE_2 G^p_c(E_2) \int dE_3 G^p_c(E_3) \int dE_{4} G^n_c(E_{4}) \delta(E_4-E_3-E_2+E_1)
     \end{eqnarray}
\end{widetext}

Here $E_1$ and $E_4$ denote the energies of the initial valence and conduction band electrons, respectively. $E_2$ and $E_3$ represent the energies of the empty conduction band states to which the electrons flow into. 
Similarly, for hole impact ionization the scattering current $I_{S}$ can also be written incorporating three energy states($E_1$, $E_2$ and $E_3$) in the valence band and one ($E_4$) in the conduction band as:

\begin{widetext} 
\centering
\begin{equation}\label{Isc_2band_hole}
I_{S} = \int dE_2 Tr\Sigma_S^{out}(E_2)G^p_v(E_2) +  \int dE_3 Tr\Sigma_S^{out}(E_3)G^p_v(E_3)- 
 \int dE_4 Tr\Sigma_S^{in}(E_4)G^n_c(E_4)- \int dE_1 Tr\Sigma_S^{in}(E_1)G^n_v(E_1) 
\end{equation}
\end{widetext}
where
\begin{widetext}
\centering
\begin{eqnarray}
\label{sig_2band_hole}
 \Sigma^{out}_S(E_{2,3} < E_{vL}) &=  D \otimes \int dE_1 G^n_v(E_1) \int dE_{3,2} G^p_v(E_{3,2}) \int dE_4 G^n_c(E_4)   \delta(E_4-E_3-E_2+E_1)\\  \nonumber  
  \Sigma^{in}_S(E_4 > E_{cR}) &= D \otimes \int dE_1 G^n_v(E_1) \int dE_2 G^p_v(E_2) \int dE_3 G^p_v(E_3) \delta(E_4-E_3-E_2+E_1)\\ \nonumber    
   \Sigma^{in}_S(E_1 < E_{vL}) &=  D \otimes \int dE_2 G^p_v(E_2) \int dE_3 G^p_v(E_3) \int dE_4 G^n_c(E_4) \delta(E_4-E_3-E_2+E_1) 
\end{eqnarray}
\end{widetext}

Note that since $E_c(x)$ and $E_v(x) $ vary with position, we greatly simplify the energy arguments above as being limited by the highest conduction band edge at the right $E_{cR}$ and the lowest valence band edge at the left $E_{vL}$. {A more accurate approach would be to introduce the $\Theta$ matrices. For instance, considering $\Sigma^{out}(E)$ for electron ionization, we could break up the energy integral into two parts separated by $\Theta(E-[E_c])$ and $\Theta([E_v]-E)$, and for each insert the right functions in the integrals,  replacing $\int dE_2G^p_c(E_2)$ with $\int dE_2 G^p(E_2)\otimes \Theta(E_2-[E_c])$, and so on.}
\\\\
The main point of the exercise above is to verify by simple substitution that the self-energies we are defining above indeed ensure that $I_{S} = 0$ in both cases. 

\subsection{The  deformation potential $D$}
In order to conserve the momentum, the $D$ matrix must be chosen appropriately \cite{golizadeh2007nonequilibrium}.
Ideally $D$ should be calculated as a fourth rank tensor, $D_{ijkl} = \langle U_{ik}U^*_{jl}\rangle$, where $U_{ij} \approx q^2e^{-\kappa|r_{ij}|}/|r|_{ij}$ in a Debye approximation, with the screening parameter $\kappa$ computed separately, for instance, by a Poisson solver.
When the non-locality in the underlying random potential $U_{ij}$ is small and well-correlated throughout the channel in real space, \textit{i.e.}, having the same value at all points of the matrix $D$, the momentum is conserved (Fourier transform of $D$ into momentum space is a delta function ensuring there is no momentum loss). The equation of $D$ is given below and in this study we consider $D_0$ to be an adjustable parameter. The physics of $D_0$ is a bilinear thermal average of the screened Coulomb potential that mediates the electron-electron collision process.  

\begin{equation}
D=D_0 \left(\begin{array}{cccccccc}
1 &  1  & 1   & 1 & 1 & 1 & \cdots & \cdots\\
1 &  1  & 1   & 1 & 1 & 1 & \cdots & \cdots\\
 \cdots & \cdots & \cdots & \cdots & \cdots & \cdots & \cdots & \cdots\\
 \cdots & \cdots & \cdots & \cdots & \cdots & \cdots & \cdots & \cdots\\
 \cdots & \cdots & 1 &  1  & 1   & 1 & 1 & 1 \\
 \cdots & \cdots & 1 &  1  & 1   & 1 & 1 & 1
\end{array}\right)
\label{eqd}
\end{equation}
Since all entries are unity, the element by element multiplication $\otimes$ ends up being just a scalar multiplication with $D_0$. 

However, for momentum-breaking processes this can be more complicated and the full $\otimes$ operation maybe necessary. A possible way to do this would be to start with phenomenological expressions \cite{Lundstrom_2000} for scattering in momentum space $U_{\vec{k}\vec{k}^\prime}$, then inverse transform along the transport direction $z$ to get a block tridiagonal matrix \cite{stovneng1993multiband} 
of the form $[U]_{\vec{k}_\perp\vec{k}^\prime_\perp}$ that still depends on transverse momenta, and then take their bilinear thermal average $\langle UU\rangle$ over electron/phonon distributions to get $D$. Typical $D$s, assuming translational invariance in the transverse $(x,y)$ directions, would then look like $D_{1234} \approx M_0^2e^{\displaystyle -iQ_z(z_3-z_4)-[(z_1-z_3)^2+(z_2-z_4)^2]/2\sigma_z^2}$, where $Q$ is the lattice momentum, $M_0$ is the strength of the scattering squared, $\sigma_z$ is the momentum scattering mean free path that denotes the spread of the diagonal elements in Eq.~\ref{eqd} into the off-diagonal space, Eq.~\ref{eqd} emerging in the limiting case where $\sigma_z \rightarrow \infty$, $Q = 0$ in the continuum approximation, and we have decoherence without momentum scattering. 
\\\\
Translating the scattering matrices into their NEGF matrix equivalents is an exercise in itself needing separate validation. 
We leave that exercise for a future publication, but our formalism for APD quantum kinetics is agnostic of those details.
 \begin{figure}[b]
    \centering
\includegraphics[width=0.4\textwidth]{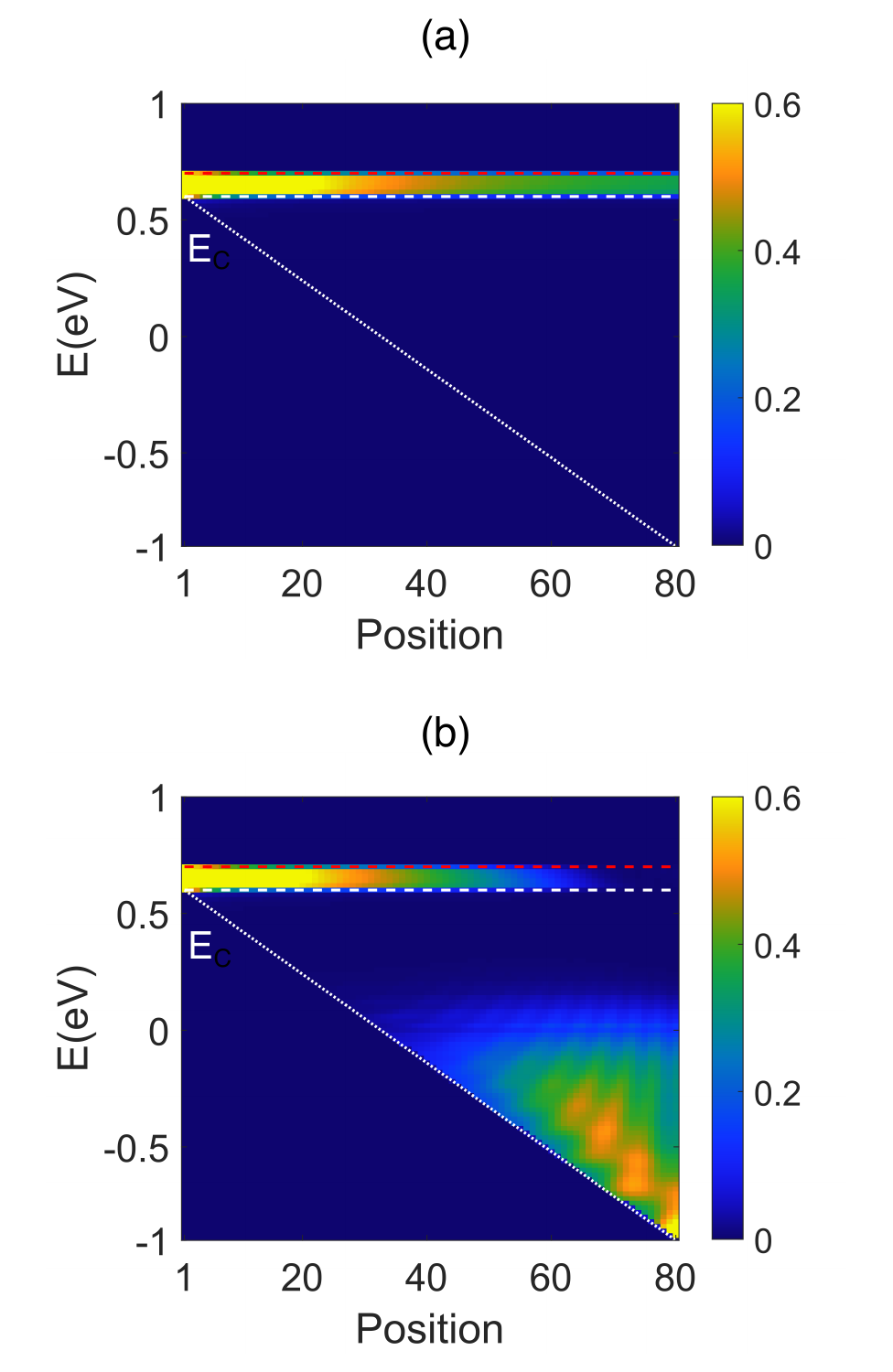}
\caption{Energy and position-resolved electron concentration in the conduction band: (a) before impact ionization and (b) after impact ionization.}
\label{fig:Impact_Ionization_LDOS}
\end{figure}

\begin{figure}[t]
    \centering
    \includegraphics[width=0.8\linewidth]{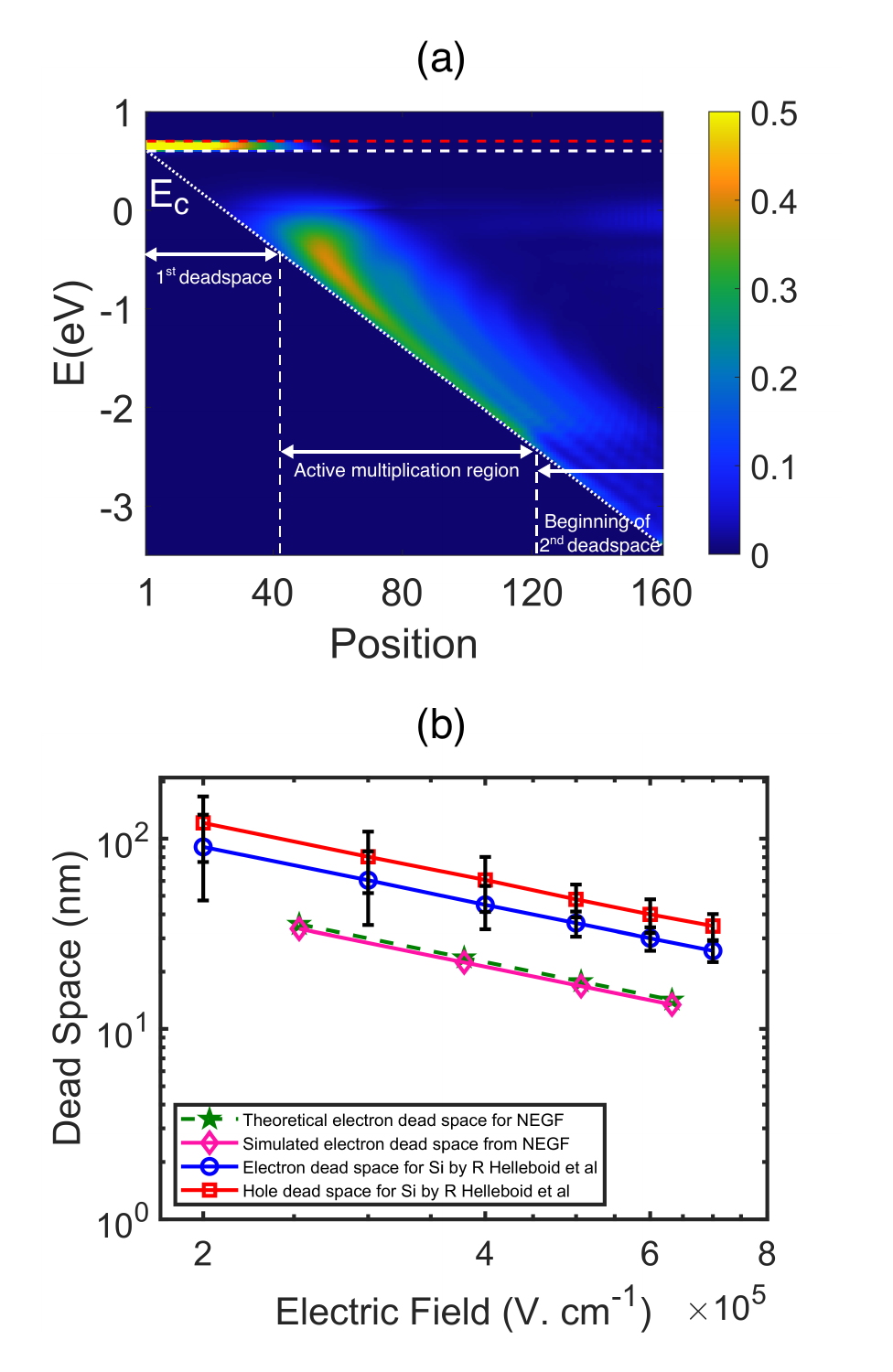}
    \caption{(a) Longer dimer chain (160 atoms) under a significantly higher electric field (4V) gives rise to multiple dead spaces in the conduction band (b) Dead space length reduces as the applied electric field increases.}
\label{fig:Multiple_deadspace}
\end{figure}

\begin{figure*} 
\includegraphics[width=0.9\linewidth]{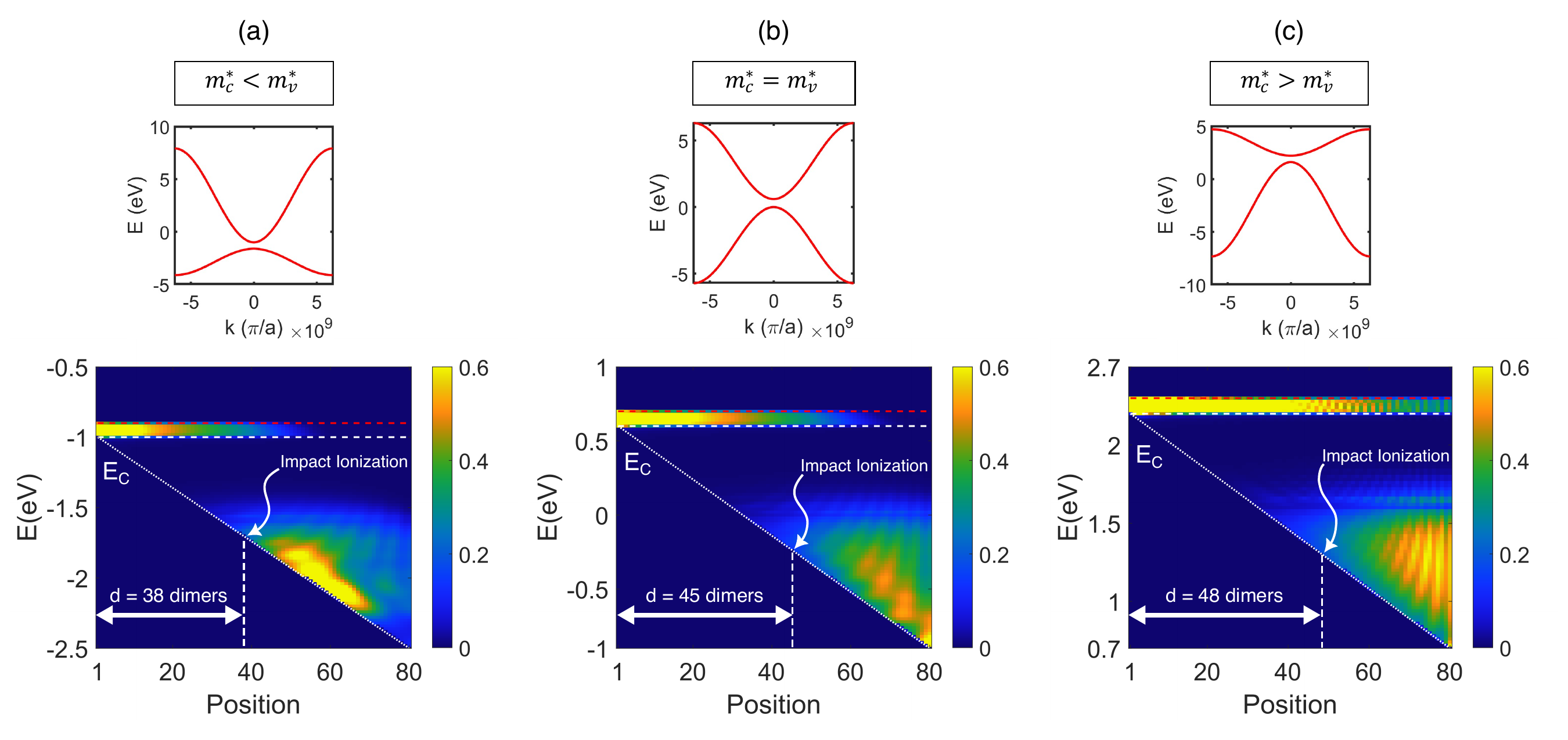}
\caption{Effect of different threshold energy on impact ionization initialization by varying effective masses (a) $m_c^*<m_v^*$ (b) $m_c^*=m_v^*$ and (c) $m_c^*>m_v^*$. There is a pronounced increase in threshold energy and dead space in dimer units.}
\label{fig:Impact_Ionization_LDOS_all}
\end{figure*}

\subsection{Results on impact ionization}
{Fig.~\ref{fig:Impact_Ionization_LDOS}(a) shows the energy resolved electron concentration $n(x,E) = [G^n(E)]_{x,x}$ in the conduction band for each dimer atom position across the simulated device {under ballistic conditions}, with an added energy filter that determines if $E$ exceeds $E_c$ at that diagonal position point. Since there is no scattering here, the window between the Fermi level $E_{F1}$ and conduction band $E_{CL}$ at the left edge depicts the ballistic transportation of the injected electrons from left to right due to the applied electric field. Fig. \ref{fig:Impact_Ionization_LDOS}(b) extends this to impact ionization, and shows that there is an accumulation of charge carriers at the right edge of the conduction band because of electron-electron scattering leading to impact ionization. These extra generated electrons will give a rise to a jump in total electron count, eventually resulting in a multiplication gain $\langle M\rangle$.} {Note also that we are simulating a short section of the superlattice, roughly 1.5 V across 80 dimers ($\sim$ 8 nm), which gives us an applied electric field ${\cal{E}} \sim 2 \times 10^6$ V/cm.}
{For practical reasons, we simulated a small segment of the APD with our dimer chain, applying a correspondingly small voltage to keep the electric field consistent with experimental estimates. To witness the impact of multiple dead spaces along the channel, we also considered a longer chain of 160 atoms with a higher bias of 4 volts (Fig. \ref{fig:Multiple_deadspace}(a)). Aside from a first impact ionization around 40 atoms, we see the smeared edges of a second dead-space around 120 atoms. Note that the only sources of smearing in our calculations are the Fermi tails of the contact electron distributions (our model at this point has no disorder or phonon scattering). This gives us the average electron distribution and current gain $\langle M\rangle$. Extracting noise from our calculation will require the NEGF-based computation of current variance, which we leave for future publications.} 

Prominent in these plots is the dead-space, the distance a carrier needs to travel on average before losing its kinetic energy and momentum to impact ionization. 
\begin{equation} \label{Deadspace_eqn}
 d_{e,h}= { E_{TH}^{e,h} }/{ q{\cal{E}}} 
\end{equation}
where $d_{e,h}$ is the dead space length, ${\cal{E}}$ is the applied electric field, and $E_{TH}^{e,h}$ is the ionization threshold energy for electron and hole injection respectively (Eq.~\ref{eth}).
The dead space depends on the local electric field ${\cal{E}}$ \cite{helleboid2022fokker}.\\

Fig. \ref{fig:Multiple_deadspace}(b) shows that the extracted electron dead space length  decreases linearly with electric field, consistent with experimental data. Our values are a little off from the experimental values, which can be attributed to the simplicity (currently 1-D) model of our impact ionization,  and the toy parameters $t_{1,4}$ with corresponding band-gap and masses that are not selected to resemble any specific material. It is also worth observing that a lot of physics can be hidden in the local electric field ${\cal{E}}$, such as the preponderance of screening with the proliferation of charges down the channel. A proper treatment of that will require including Poisson's equation self-consistently with the Green's function treatment of non-equilibrium charge distribution \cite{damle}.

% \begin{figure*}
% \includegraphics[width=0.7\linewidth]{All_III-V_dimer_impact.png}
% \caption{Visualization of impact ionization happening in the conduction band in every III-V binary alloy materials. Very narrow band-gap materials such as InSb also shows the instance of B2B tunneling inherently on top of the impact ionization.}
% \label{fig:Impact_Ionization_all_III-V}
% \end{figure*}

% \begin{figure} 
% \includegraphics[width=\columnwidth]{Impact_LDOS_all_mass.pdf}
% \caption{Effect of different threshold energy on impact ionization initialization by varying effective masses (a) $m_c^*<m_v^*$ (b) $m_c^*=m_v^*$ and (c) $m_c^*>m_v^*$. There is a pronounced increase in threshold energy and dead space in dimer units.}
% \label{fig:Impact_Ionization_LDOS_all}
% \end{figure}

Fig. \ref{fig:Impact_Ionization_LDOS_all} shows the impact of band effective masses on where the impact ionization initiates. Changing the effective masses by varying the dimer parameter $t_4$ changes the threshold energy for impact ionization, which for parabolic bands can be written as 
\begin{equation}
    E_{TH}^{e,h}=[(2\mu+1)/(\mu+1)]E_G
    , ~~~ \mu = m^*_{c,v}/m^*_{v,c}
\label{eth}
\end{equation}
For example, in Fig. \ref{fig:Impact_Ionization_LDOS_all}(a), where $m_c^* < m_v^*$, the threshold becomes $E_{TH}^e < 1.5~E_G$ and it impact ionizes relatively earlier than the case of $E_{TH}^e = 1.5~E_G$ ($m_c^* = m_v^*$) in Fig. \ref{fig:Impact_Ionization_LDOS_all}(b). For $m_c^* > m_v^*$, we get $E_{TH}^e > 1.5~E_G$ which delays the impact ionization later than Fig. \ref{fig:Impact_Ionization_LDOS_all}(b). Thus the dead space distance is related to this threshold energy by the applied electric field, $\cal{E}$.

\begin{figure}[b]
\centering
\includegraphics[width=0.8\columnwidth]{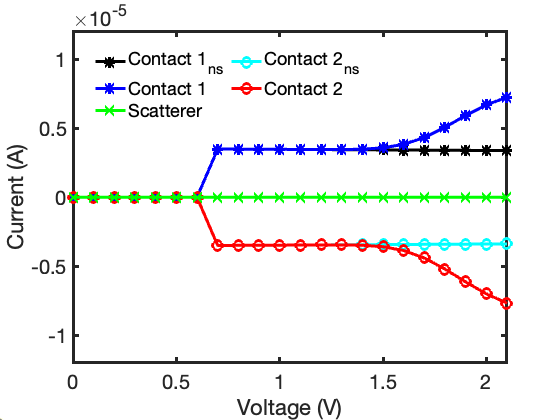}
\caption[Current vs. voltage characteristics of a 1D semiconductor with impact ionization.]{Current vs. voltage characteristics of a 1D semiconductor with impact ionization. Compare with Fig.~\ref{fig:four_level_IV}.{A ballistic current is initiated at $~0.6$ volt when the inverted n-i-p structure in Fig.~\ref{fig:1D_device} goes slightly above flatband conditions. After a further threshold voltage $V_{TH} = 0.9$ V, set by Eq.~\ref{eth} for a bandgap of $0.6$ eV and equal masses ($\mu = 1$), the current starts jumping exponentially because of impact ionization. The slight roll-off in current at higher voltages is an artifact of charge throttling in 1-D (Fig.~\ref{fig:DOS_sliding}), where at higher energies, the incoming electrons encounter progressively lower densities of states in the drain. }}\label{fig:IV_dimerchain}
\end{figure}

To extract the threshold voltage and quantify these effects, we plot the simulated terminal current vs. voltage of a 1D semiconducting dimer chain with a length of 80 dimers (Fig. \ref{fig:IV_dimerchain}). We set $D_0=5$, the temperature $T=3~K$, band gap $E_G=0.6~eV$, and electron at 0.1 eV above the left contact Fermi energy. For this plot, conduction and valence band effective masses are considered to be equal ($t_4=0$). We calculate scattering current only for electrons, since {primary carriers are electrons photo-excited into the conduction band, and secondary ionizations are less consequential to the overall gain (although they matter more for current variance, ie, excess noise).} The ballistic currents (non-scattering, marked `ns', meaning without impact ionization), labeled Contacts ${1_{ns}}$ and ${2_{ns}}$, jump  {after an initial voltage needed to reach flatband conditions from the inverted n-i-p structure (Fig.~\ref{fig:1D_device}), and then saturates, staying equal and opposite for the two contacts. 
Under impact ionization, the terminal currents (labeled Contact 1 and 2) increase after reaching an additional threshold voltage, which for an equal mass system ($\mu = 1$ in section \ref{II1D}), is $E_{TH} = 1.5E_G = 0.9$ eV. 
}

\begin{figure}[b]
    \centering
    \includegraphics[width=0.5\textwidth]{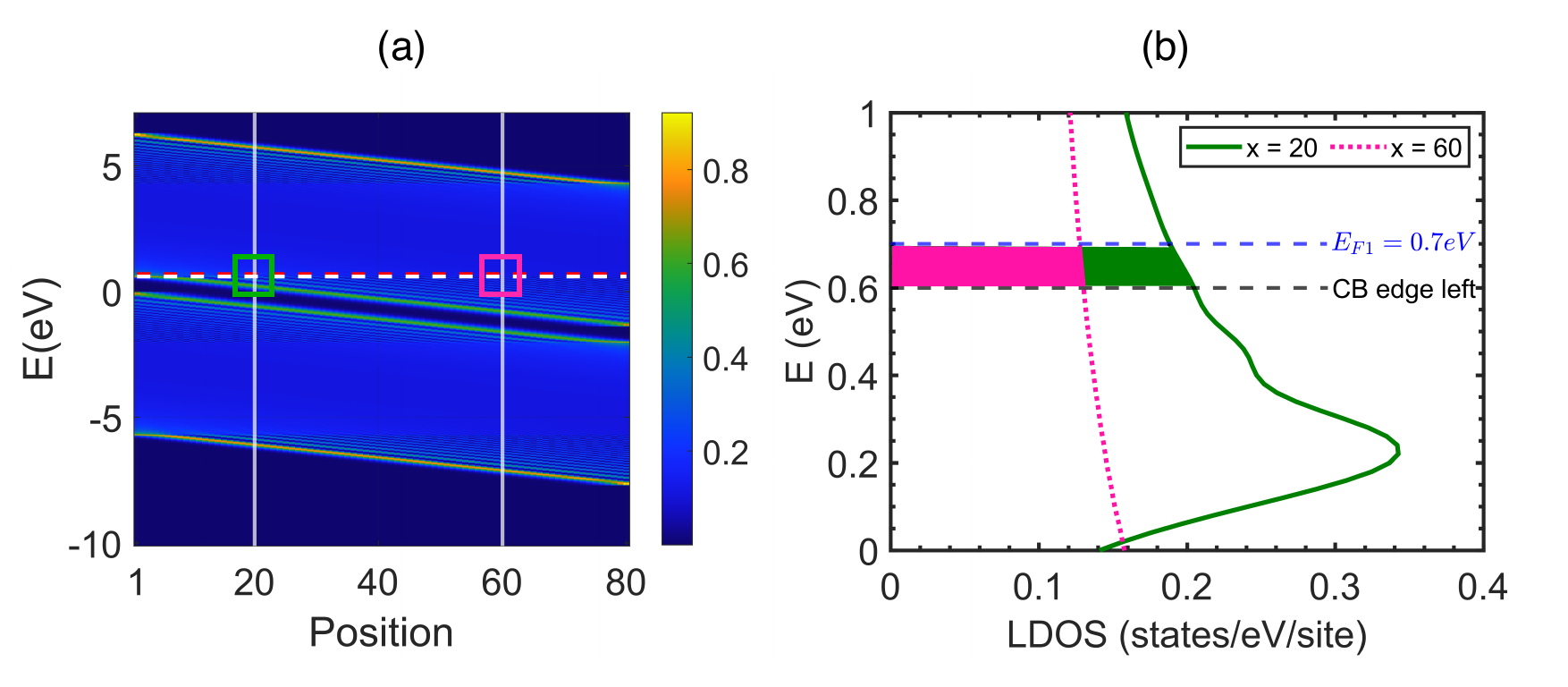}
    \caption{(a) LDOS plot of the simulated 1D semiconductor dimer chain structure (b) Going from left ($20\textsuperscript{th}$ atom) to right ($60\textsuperscript{th}$ atom) shows less electron occupation area under the DOS curves}
    \label{fig:DOS_sliding}
\end{figure}

\begin{figure}[t]
    \centering
    \includegraphics[width=0.5\textwidth]{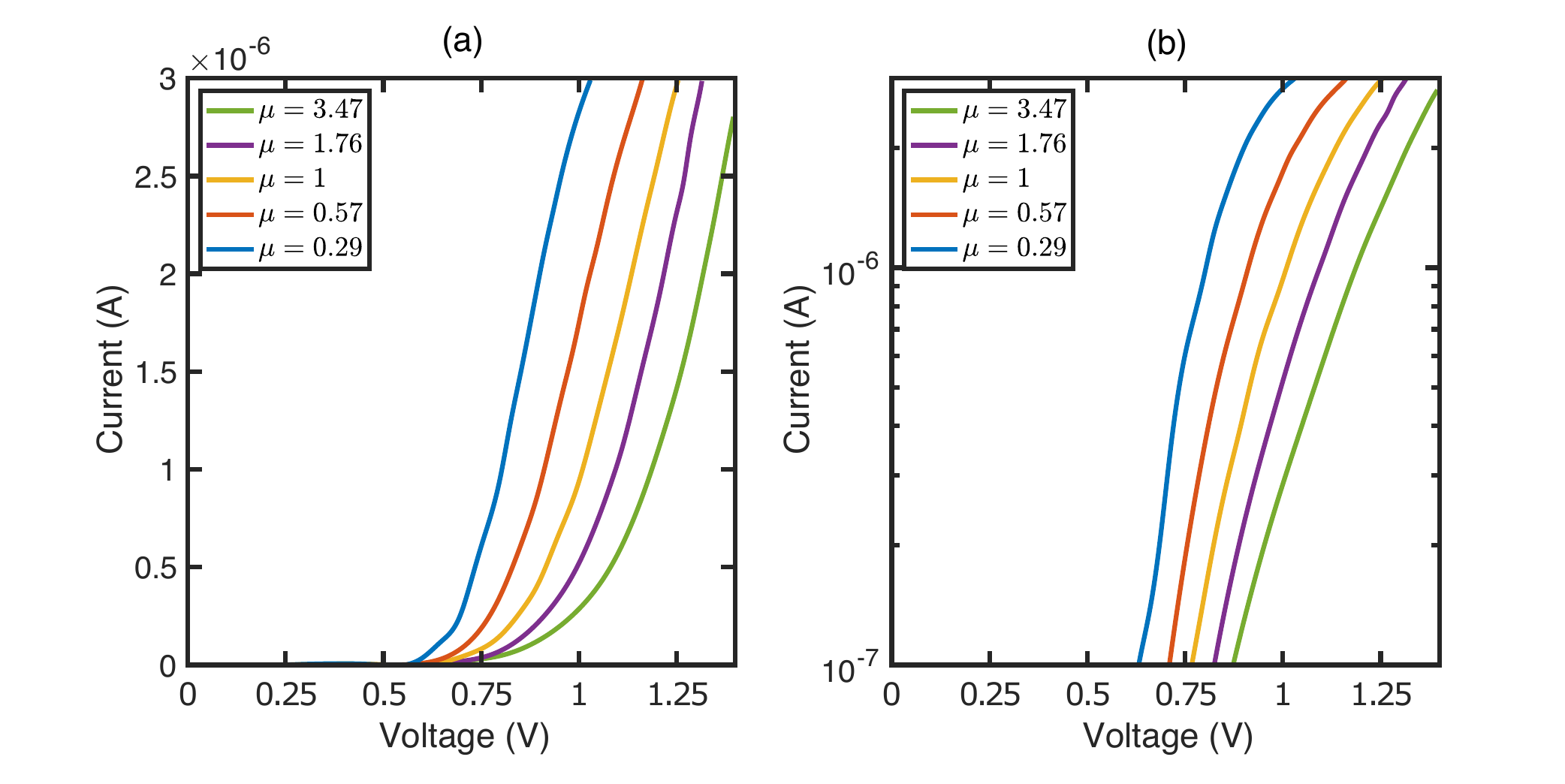}
    \caption{Impact Ionization current vs. voltage characteristics for different mass ratios $\mu$: (a) linear scale and (b) semi-logarithmic scale.}
    \label{fig:mu_threhsold}
\end{figure}

\textcolor{black}{There is a high bias roll-off of the exponentially initiated impact ionization current. A potential origin of this roll-off is a peculiarity of our 1-D model.  Fig.\ref{fig:DOS_sliding} (a) shows a colorplot of the local density of states (LDOS) $A(x,x,E)$ of our simulated 1D semiconductor dimer chain structure. If we observe from left to right along the atomic position, we can see a reduction in the density of states, since the 1-D DOS decreases with energy as $\sim 1/\sqrt{|E-E_{c,v}(x)|}$ relative to the spatially sliding band edges. Consider the area of the electron injection region =between the conduction band edge at the left and the Fermi level $E_{F1}$. In Fig.\ref{fig:DOS_sliding} (b), we see that the area under the LDOS at $60\textsuperscript{th}$ atom is less than that at the $20\textsuperscript{th}$. The reduction of states to inject into `throttles' the injected charges that are forced to reflect, countering the exponential increase in avalanche current.} Note that the lower dimension does not affect the impact ionization itself, which is dominated by electron heating by the large electric field.
 %{\textcolor{blue}{To compute the terminal currents with impact ionization in a 1D semiconductor, we consider the setup as shown in Fig. \ref{fig:1D_device}. The quasi-Fermi level of the left contact $E_{F1}$ is set above the conduction band edge on the left which essentially makes it an NIP photo diode structure instead of the traditional PIN structure. However, this facilitates the injection of electrons into the channel, mimicking the process of electron injection due to photon absorption. Thus it can simplify the quantum transport simulation of our model. The quasi-Fermi level of the right contact $E_{F2}$ is fixed to the right valence band edge. The voltage $V$ is the applied potential across the channel.} 

% \begin{figure}[t]
%     \centering
%     \includegraphics[width=0.5\textwidth]{I-V_threshold.pdf}
%     \caption{Impact Ionization current vs. voltage characteristics for different mass ratios $\mu$: (a) linear scale and (b) semi-logarithmic scale.}
%     \label{fig:mu_threhsold}
% \end{figure}

\begin{figure}[b]
\centering
\includegraphics[width=0.4\textwidth]{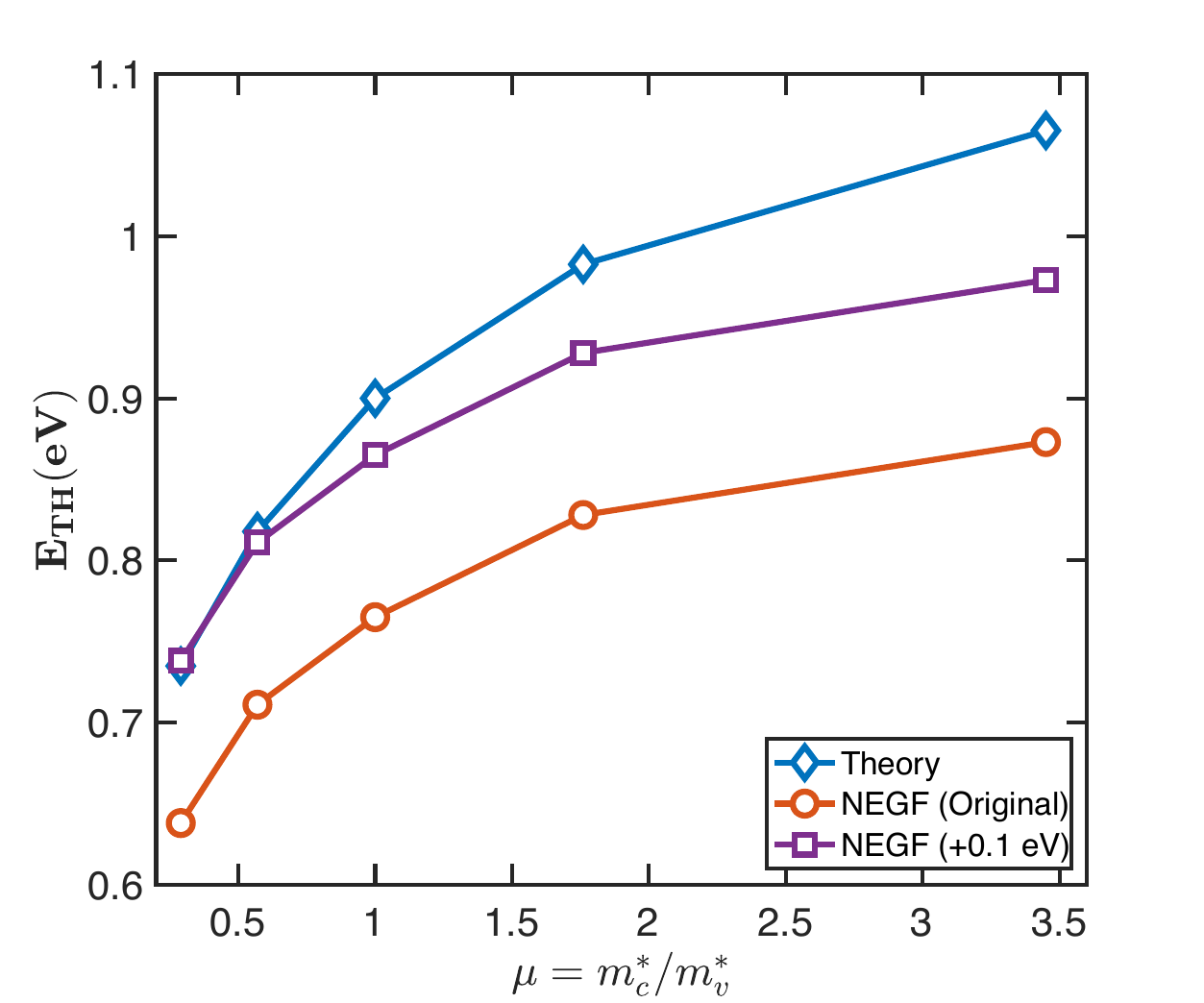}
\caption{\textcolor{black}{Ionization threshold energy plot showing results from analytical theory (Eq.~\ref{eth}), the raw result from NEGF, and finally NEGF adjusted for a 0.1 eV overdrive, since the electrons are injected above the band-edge. At low $\mu$ (infinitely heavy holes) we can safely ignore secondary hole ionization, but as $\mu$ increases, the hot holes impact ionize and generate additional cold electrons, increasing the electron threshold, which our unipolar calculation under-estimates.}}\label{fig:eth_adjusted}
\end{figure}

Fig. \ref{fig:mu_threhsold} depicts the impact ionization current (total terminal current minus ballistic current) vs. voltage characteristics for different effective mass ratios $\mu$, where $\mu=m_c^*/m_v^*$ for electron injection. We observe that the turn on voltage for the impact ionization increases with increasing $\mu$. The impact ionization current increases with voltage because carriers with lower kinetic energy can impact ionize at higher voltages. {We can extract the threshold voltages from a semilog plot.} The extracted threshold energy as a function of $\mu$ is shown in Fig. \ref{fig:eth_adjusted} for a semiconductor with $E_G=0.6~eV$. $E_{TH}^{e}$ approaches a value of $2E_G$ as $\mu\rightarrow \infty$ and goes toward $E_G$ as $\mu\rightarrow 0$. The NEGF threshold energy exhibits the same trend as the ideal threshold energy (Eq.~\ref{eth}).  The offset between the two threshold energies can be attributed to the extra kinetic energy of the injected electrons due to the quasi-Fermi level of the left contact $E_{F1}$ being slightly above $E_C$ by about 0.1 eV. In Fig. \ref{fig:eth_adjusted} the ionization threshold, increased for the energy shift, is seen to  catch up with the ideal curve.

\begin{figure}[ht!]
\includegraphics[width=0.5\textwidth]{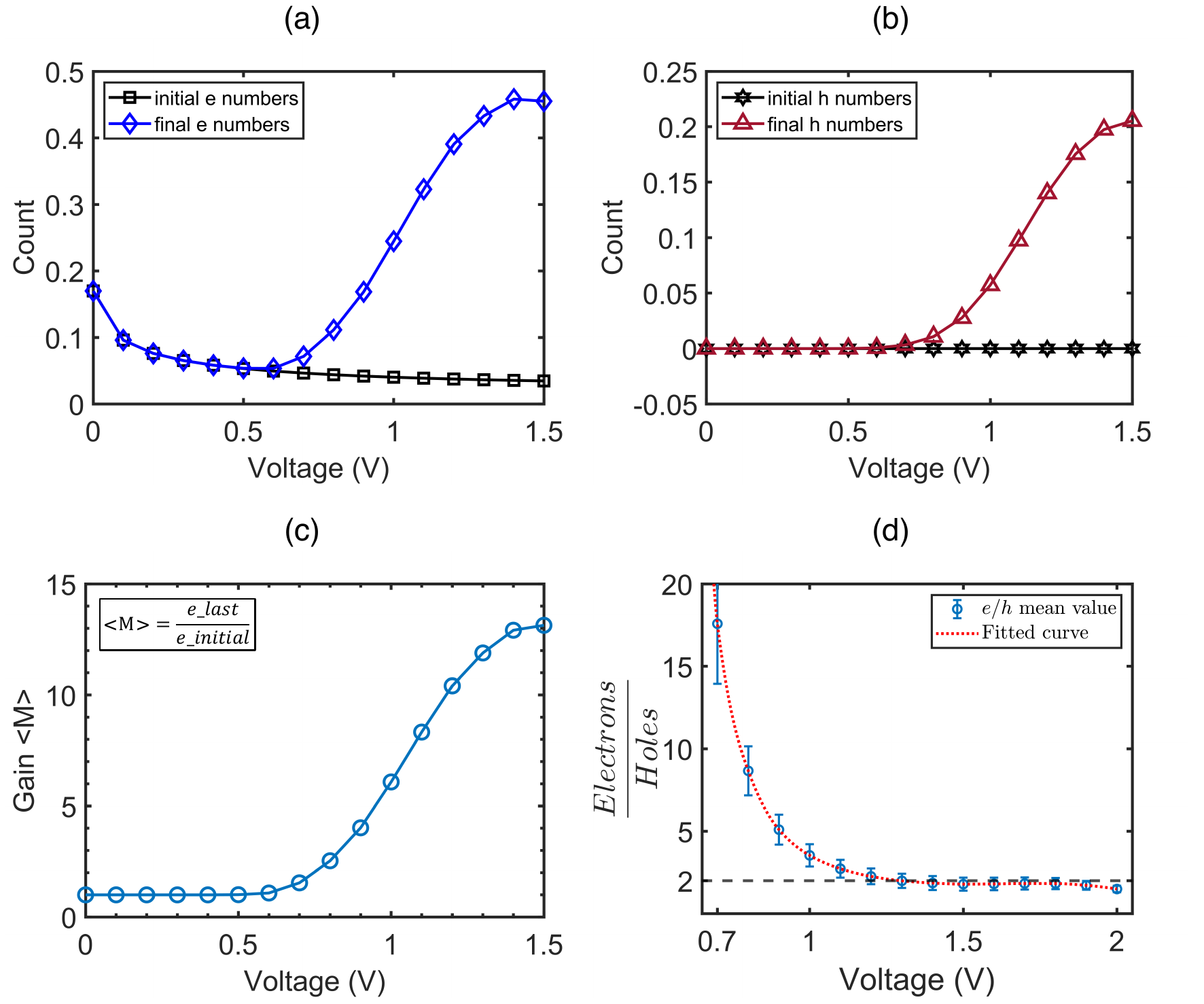}
\caption{\label{fig:APD_Gain} (a) Electron count and (b) hole count are going up after the electron scattering event, which indicates the targeted impact ionization event. The initial drop is a consequence of charge throttling specific to 1-D ballistic transport. (c) Taking ratio between electron counts before and after gives the multiplication gain, M (d) Electrons and holes keep a ratio of 2:1 after the scattering event, plotted over a range of $D_0$ values.}
\end{figure}

Let us verify one more signature of impact ionization.
Fig.\ref{fig:APD_Gain}(a) shows the number of electrons at the right end of the device, without and with inelastic scattering, and (b) number of holes at the left end, as a function of applied reverse bias varying from 0V to 1.5V. We see that the number of electrons doubled at 0.78 V. Both charge distributions show an exponential rise at the threshold voltage $\sim 0.78$ V. The initial drop in electron number before impact ionization is an artifact of charge throttling in 1-D, as described earlier (Fig.~\ref{fig:DOS_sliding}). Fig.~\ref{fig:APD_Gain}(c) shows the charge gain $\langle M\rangle$ across the multiplication region, obtained by subtracting the ballistic charge distribution. Since we ignore secondary ionization of holes, all holes arise only through primary ionization of electrons rather than direct injection, meaning their count stays close to zero until impact ionization happens. More promisingly, while the ratio of electron to hole count starts off near infinity (no holes) for low bias, it quickly saturates closer to two, as each primary electron ionization ends up with two electrons in the conduction band and one hole in the valence band (Fig.~\ref{fig:APD_Gain}(d)). The results are shown here for a distribution of $D_0$ values in the deformation potential.\\

\begin{figure*}
\includegraphics[width=0.9\linewidth]{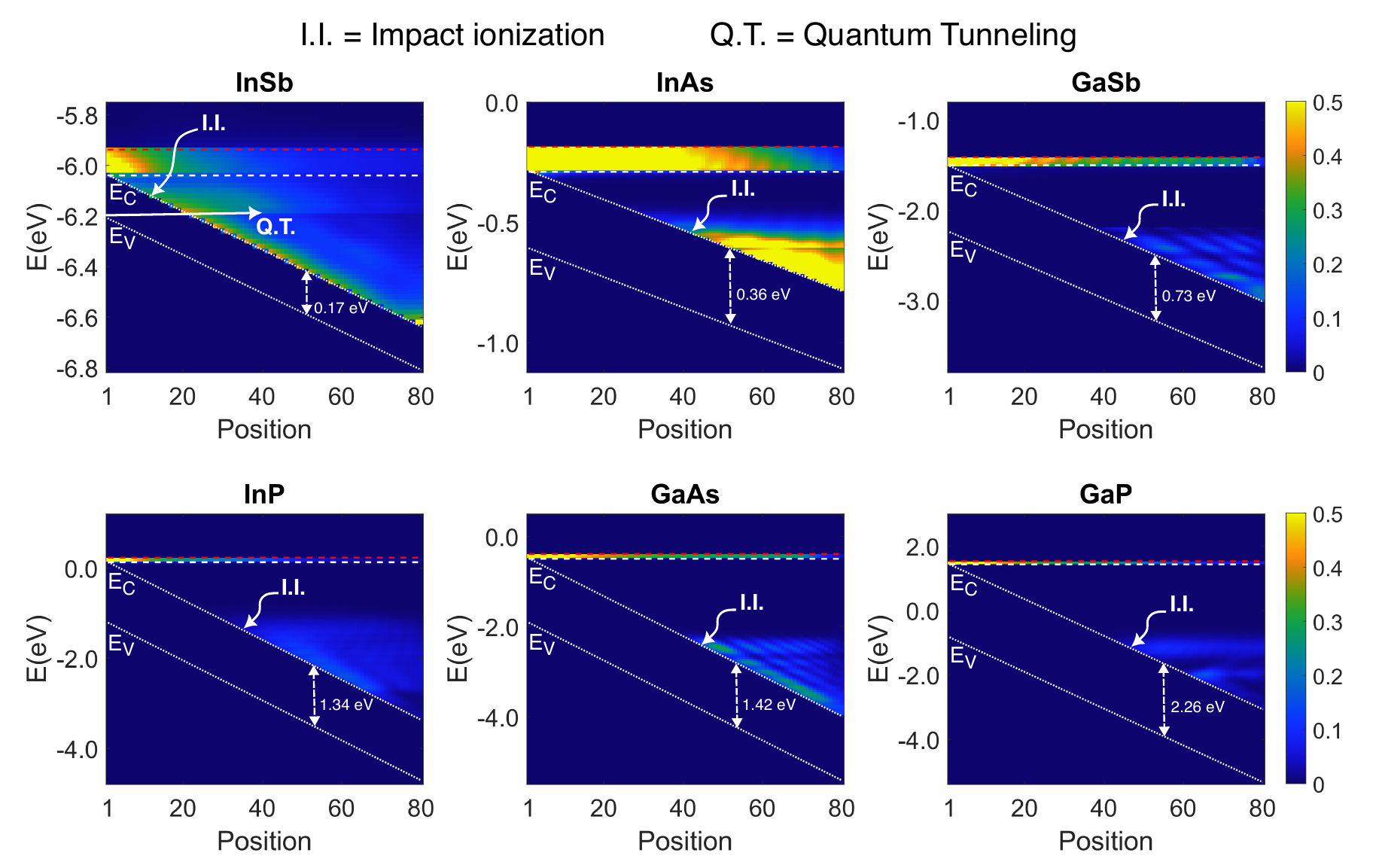}
\caption{\textcolor{black}{Visualization of impact ionization happening in the conduction band in commonly used III-V binary alloy materials. Very narrow band-gap materials such as InSb also shows the instance of B2B tunneling inherently on top of the impact ionization, which contributes to their dark current.}}
\label{fig:Impact_Ionization_all_III-V}
\end{figure*}

\begin{table*}[htb]
\centering
\caption{\textcolor{black}{Tight-binding parameters including onsite energies ($\varepsilon_1$, $\varepsilon_2$) and hopping coefficients ($t_1$ to $t_4$), chosen to reproduce the target electronic properties of each material: band gap ($E_g$), conduction band effective mass ($m^*_c$), and valence band effective mass ($m^*_v$) \cite{vurgaftman2001band}. The materials are arranged in order of increasing target band gap.}}
\label{tab:material-params}
\begin{tabular}{l@{\hskip 12pt}c@{\hskip 12pt}c@{\hskip 12pt}c@{\hskip 12pt}c@{\hskip 12pt}c@{\hskip 12pt}c@{\hskip 12pt}c@{\hskip 12pt}c@{\hskip 12pt}c}
\hline \hline
\textbf{Material} & $\varepsilon_1$ & $\varepsilon_2$ & $t_1$ & $t_2$ & $t_3$ & $t_4$ & $E_g$ (eV) & $m^*_c$ ($m_0$) & $m^*_v$ ($m_0$) \\
\hline
InSb  & 0.23002  & 0.06074   & 11.8972 & -5.61254 & -6.27705 & 3.1349   & 0.17 & 0.014 & 0.43 \\
InAs  & 0.24603  & -0.07655  & 25.9306 & -13.8245 & -12.186  & 2.12937  & 0.36 & 0.023 & 0.41 \\
GaSb  & 0.73067  & 0.00154   & 2.49749 & -0.74508 & -1.77026 & 1.12095  & 0.73 & 0.041 & 0.40 \\
InP   & 1.31116  & 0.00278   & 5.12581 & -1.96381 & -3.30669 & 0.59908  & 1.34 & 0.080 & 0.60 \\
GaAs  & 0.68041  & 0.00152   & 1.83054 & -0.36606 & -2.08808 & 0.77151  & 1.42 & 0.067 & 0.50 \\
GaP   & 2.2518   & -0.00084  & 2.93461 & -0.91247 & -1.93102 & 0.41227  & 2.26 & 0.130 & 0.79 \\
\hline \hline
\end{tabular}
\end{table*}

It is worth clarifying that since our mass ratio and thus the ratio $k$ of hole to electron ionization rates is non-zero, we expect to see secondary hole ionization and a $k$-dependent ratio $(2+k)/(1+2k)$ of charge gain. Capturing this effect will require extending self-consistency beyond just primary electron ionization, changing the integral limits in Eqs. \ref{Isc_2band} and \ref{sig_2band}. It is possible to extend this matrix-based quantum mechanical treatment of impact ionization to devices with complicated material band structures and quantum effects like tunneling across minigaps \cite{apd_inequality}. 
\subsection{Application to III-V materials}
Table \ref{tab:material-params} and Fig.~\ref{fig:Impact_Ionization_all_III-V} show the extension of our dimer model to a variety of III-V materials. We can fit both conduction and valence band effective masses and the bandgaps with the simplified dimer model, although it is not designed to capture, in its present form, any anisotropy, indirect band-gaps or non-parabolicity (for which a proper tight-binding based APD solver is needed). We see the onset of impact ionization and their dead-space dependence on band-gap. InAs gives a large amount of impact ionization because its bandgap and effective masses are small. InSb on the other hand, shows a clear onset of tunneling (an added slice in the colormap around $E = -6.2$V), which increases its dark current.

\section*{Conclusion}
In this paper, we introduced a matrix-based quantum transport model for impact ionization using the Non-Equilibrium Green's Function formalism, with a self-energy based on multiparticle collisions. This can be compared with GW approximations that avoid the explicit many-particle interaction with a resummed bubble diagram by rolling two of the $G$ products into an effective screening kernel $W$ \cite{Luisier_GW_iedm,Thygesen_GW_prb}. We illustrate our approach with a minimal four-state model, and a 1-D model dimer based semiconductor. The model exhibits behavior expected of such a material - strict vanishing of scattering current, exponential increase in terminal current at a predictable threshold voltage, dead-spaces for ionization that vary with mass and field in expected ways, increase in tunneling at some bandgaps and low masses, and a predictable charge gain at each collision event.  The framework lays the groundwork for complicated heterostructures with multiple folded bands, anisotropic, energy-dependent mass tensors, transverse momentum-dependent deformation potentials that are more sophisticated than Eq.~\ref{eqd}, and quantum mechanical tunneling, as captured through an atomistic matrix Hamiltonian. In future extensions, we will combine it with electron-phonon scattering self-energies, an atomistic sp3s* Hamiltonian, better treatment of convergence both across electrons and holes, and a real 3D crystal structure to avoid charge throttling. In addition, going beyond average current to calculate variance using a matricized B\"uttiker approach \cite{Ghoshbook2} will allow us thereafter to look at excess noise. 

\section*{Acknowledgment}
This work was funded by National Science Foundation grants NSF 1936016 and NSF 2430629. The authors thank Dr. John P David of University of Sheffield, and Dr. Seth R. Bank of the University of Texas-Austin for important discussions and insights. The calculations were done
using the computational resources from High-Performance
Computing systems at the University of Virginia (Rivanna)
and the Extreme Science and Engineering Discovery Environment (XSEDE), which was supported by National Science Foundation grant number ACI-1548562.

\section*{Supplementary Sections}
\begin{figure} [b]
    \centering
    \includegraphics[width=0.4\textwidth]{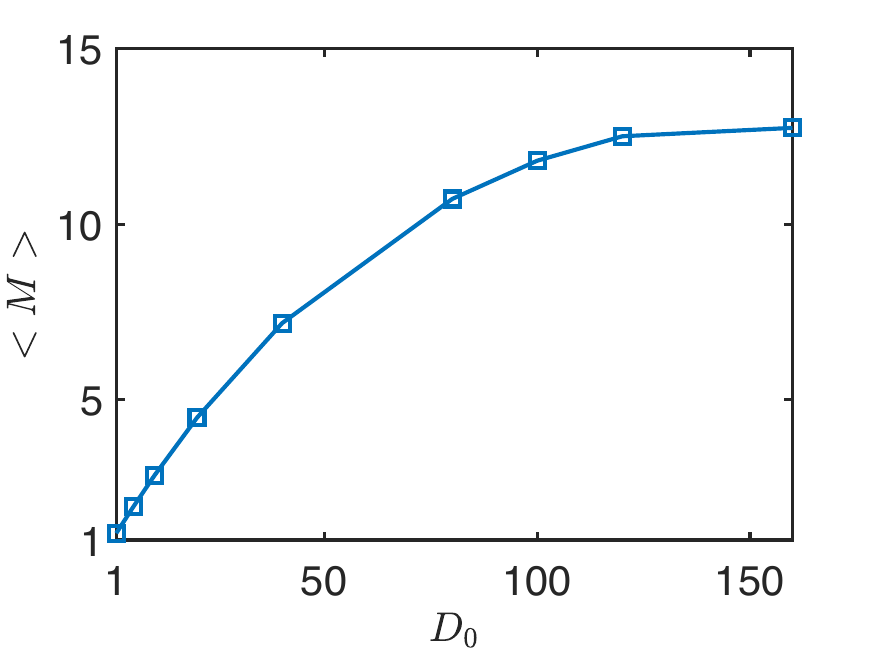}
    \caption{Increasing the electron-electron interaction matrix $D_{0}$ increases the impact ionization rate which helps to achieve relatively higher multiplication gain}
    \label{fig:D0 effect on gain}
\end{figure}

In this section, we present two technical details. As we increase the electron-electron Coulomb scattering $D_0$, we see (Fig.~\ref{fig:D0 effect on gain}) that the average gain $\langle M\rangle$ also increases until it saturates. 

\begin{figure}[t]
    \centering
    \includegraphics[width=0.4\textwidth]{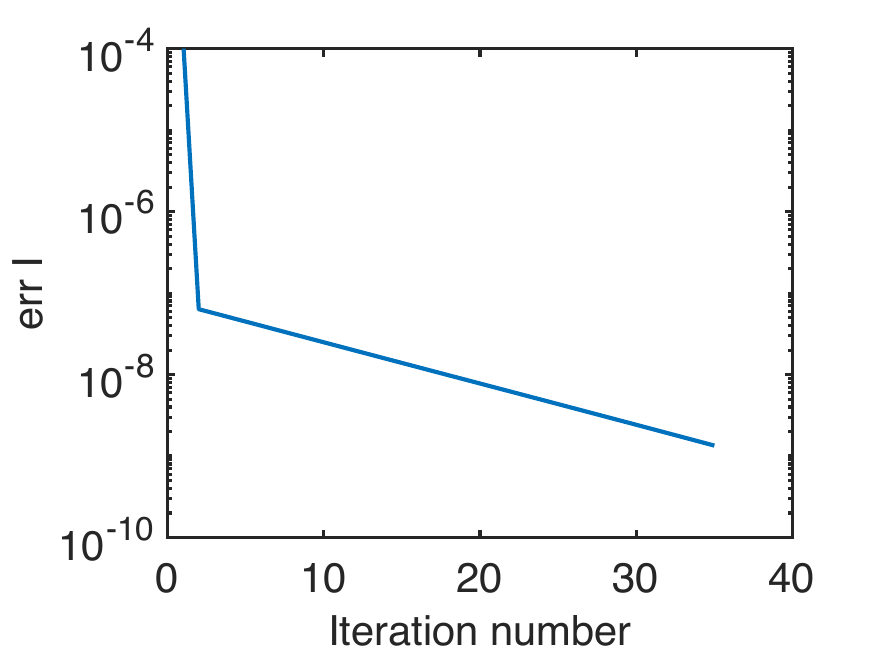}
    \caption{Convergence of the self-consistent NEGF calculation with electron-electron scattering, quantified by the current error $\text{errI}(l) = |I_{l-1} - I_l|$, where $I_l = \Delta E \sum_{k} I_1(k)$ is the total current at iteration $l$. The rapid initial decrease (from $10^{-4}$ to $10^{-7}$) within the first few iterations indicates efficient capture of dominant scattering processes, followed by a steady exponential decay reaching $10^{-9}$ by iteration 35, demonstrating the numerical stability implemented in this work.}
    \label{fig: error in I}
\end{figure}

\begin{figure}[htbp]
    \centering
    \includegraphics[width=0.4\textwidth]{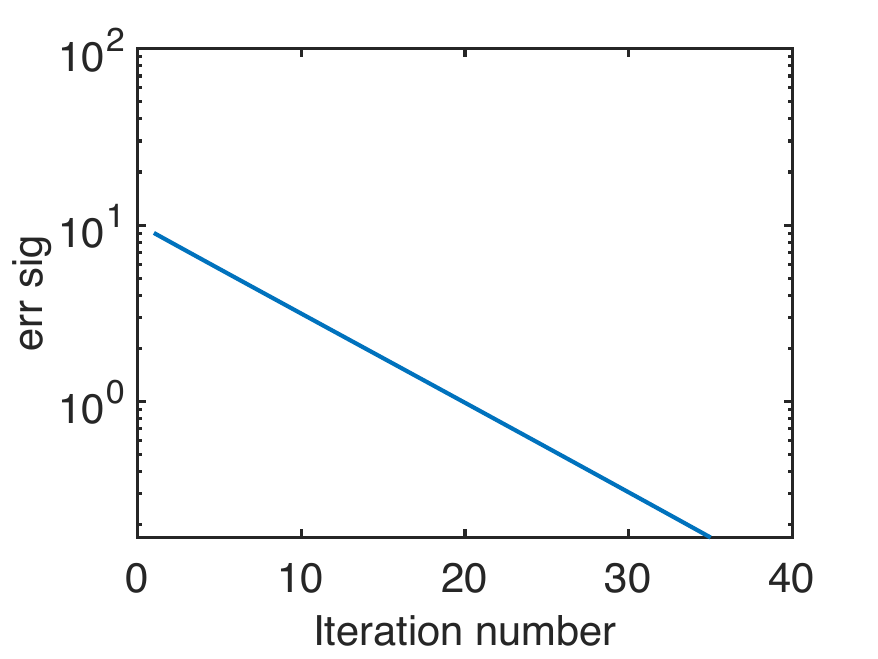}
    \caption{Convergence of the scattering self-energy matrices in the self-consistent NEGF calculation, quantified by $\text{err\_sig}(l) = |\sigma_{l-1} - \sigma_l|$, where $\sigma_l = \sum_{k} \sum_{i,j} [\Sigma_{\text{in}}(i,j,k) + \Sigma_{\text{out}}(i,j,k)]_l$ represents the sum of all matrix elements across energy points at iteration $l$. The consistent log-linear decay from $10^1$ to $10^{-1}$ over 35 iterations demonstrates robust matrix convergence, which is particularly significant as self-energy matrix convergence provides a more rigorous and fundamentally sound criterion for numerical stability than scalar quantities alone. This systematic decrease in matrix error confirms the proper implementation of energy and momentum conservation during electron-electron scattering events within the quantum transport formalism.}
    \label{fig:error in sigma}
\end{figure}

Critical to our calculation is the ability to converge our self-consistent results, since the $G^{n,p}$ correlations depend on self-energy through the Keldysh equation, while the self-energies depend on $G^{n,p}$ through their interaction term. We ignored hole ionization to simplify the calculation to unipolar effects; nonetheless, we needed to demonstrate convergence. Instead, we monitored the difference between successive iterations of the current (Fig.~\ref{fig: error in I}) and the maximum self-energy $\Sigma$ (Fig.~\ref{fig:error in sigma}).  Each error drops abruptly, indicating that the terms have converged. 

\clearpage
\bibliography{apssamp}% Produces the bibliography via BibTeX.

\end{document}